\documentclass[a4paper,11pt]{article}
\pdfoutput=1 % if your are submitting a pdflatex (i.e. if you have
\usepackage{jcappub} % for details on the use of the package, please
\usepackage[T1]{fontenc} % if needed

\usepackage{subfigure} 

\usepackage[left]{lineno}
\usepackage{amsmath}
\usepackage{graphicx}
\usepackage{multirow,multicol}

\newcommand{\mockone}{\textit{Mock 1}~}
\newcommand{\mocktwo}{\textit{Mock 2}~}
\newcommand{\mockthree}{\textit{Mock 3}~}
\newcommand{\mockfour}{\textit{Mock 4}~}

\title{Fast Mock Catalog Generation for Large Scale Structure Modeling}

\author[a,1]{T. Yapici \note{Corresponding author},}
\author[a]{Z. Brown,}
\author[a]{R. Demina,}
\author[a]{S. Benzvi}

\affiliation[a]{Department of Physics and Astronomy, University of Rochester, Rochester, NY 14627, USA}

\emailAdd{tyapici@ur.rochester.edu}
\emailAdd{zbrown5@ur.rochester.edu}
\emailAdd{regina.demina@rochester.edu}
\emailAdd{sybenzvi@pas.rochester.edu}

\date{Accepted XXX. Received YYY; in original form ZZZ}

\abstract{
To understand the universe and to interpret the cosmological parameters governing its evolution it is necessary to contrast the data from galaxy surveys with simulation. Typically it entails using  computationally expensive $N$-body simulations. Computational overhead makes it difficult to test the dependence of galaxy large scale structure on multiple cosmological parameters. In this work, we suggest a parametric model to simulate large scale structure. The new method provides a fast way to generate mock catalogs for testing multiple cosmological parameters as well as providing a test bench for code development.  
}

\begin{document}
\label{firstpage}
\maketitle

%
% this part is only temporary to make the page layout good
~\\~\\~\\

\section{Introduction}
\label{sec:intro}
Future high statistics experiments will allow us to understand the cosmological phenomena affecting the large scale structure of the universe. Increased sample sizes will demand a control over the systematic uncertainties down to sub-percent level, which requires detailed models that reproduce features observed in data. One of such features, the Baryon Acoustic Oscillation (BAO), is an imprint on galaxy distribution of density waves caused by the  fluctuations in primordial matter distribution~\cite{Peebles:1970ag}. BAO signal was observed by spectroscopic galaxy surveys, where the galaxy coordinate along the line of sight (LOS) is inferred from its redshift~\cite{Eisenstein:2005su}. The peculiar motion of the galaxies results in the distortion of its position. This effect is referred to as the Redshift-space distortions (RSD)~\cite{Kaiser:1987qv} and is a property dependent upon the growth of the universe.

To model the observed galaxy distribution, one needs to run a full simulation of the Universe evolving from the era of decoupling to present day.  This is usually done by employing $N$-body simulations, which use general relativity formalism to describe the dark matter kinematics, as well as the thermodynamics of the gas and matter. A detailed history about the N-body simulations is provided in~\cite{doi:10.1146/annurev.astro.36.1.599}. Full N-body simulation is a very CPU consuming operation, which makes it challenging to explore of effect of 
the number of possible cosmological parameters.   

The parametric model described in this paper is developed to significantly reduce the computational time, while reproducing the main features, e.g. BAO and RSD, present in the observation data, or the full N-body simulations. The paper is structured as follows: in Section~\ref{sec:model}, we explain our parametric model. In Section \ref{sec:performance} we present the tests of the performance on generated  Azimov `toy' data sets, i.e. mocks, using two point correlation function and the corresponding power spectra.

\section{\boldmath Description of the Model}
\label{sec:model}

Using our parametric model, we construct mock galaxy catalogs that account for the large scale structure of the universe given a cosmological model. A mock galaxy catalog consists of several populations of galaxies distributed over the surveyed volume satisfying an angular mask that encompasses the region of interest. These populations include:
\begin{itemize}
    \item  uniformly distributed galaxies, referred to as ``flat'' galaxies ,
    \item galaxies that carry the imprinted BAO signal,
    \item ``clumped'' galaxies that model the gravitationally bound galaxy clusters.
\end{itemize}
 All populations must follow the redshift distribution of the survey. This procedure that ensues this is discussed in Sec.~\ref{sec:acceptance}.

The distance of a given galaxy from the Earth, $r$, is calculated from its observed redshift $z$ in the comoving frame of reference given a set of cosmological parameters as:
\begin{equation} \label{eq:comovingr}
r(z) = \frac{c}{H_0} \int_0^z dz'\, \left(\Omega_M(1+z')^3+\Omega_k(1+z')^2+\Omega_\Lambda\right)^{-1/2} = \frac{c}{H_0} I(z)
\end{equation}
where $c$ is the speed of light, $H_0$ is the present day value of the Hubble constant, $\Omega_M$ is the relative matter density of the universe, $\Omega_k$ measures the curvature of space, and $\Omega_\Lambda$ is the relative density due to the cosmological constant. In the parametric mocks these parameters are defined by the user. In this paper we assume a $\Lambda$CDM+GR flat cosmology with parameters consistent with those used in the analysis of the DR9 SDSS data set \cite{Anderson:2012sa}, e.g., $H_0=70$~km~s$^{-1}$~Mpc$^{-1}$, $\Omega_M$=0.31, $\Omega_\Lambda$=0.69, and $\Omega_k$=0. 

\subsection{Generating galaxies}

\paragraph{Flat Galaxies:}
To model random distribution galaxies are distributed over the survey volume by assuming uniform distribution in right ascension and declination over the surveyed region determined by the user-defined angular mask. The distribution over the redshift also follow the user-defined redshift distribution, $P(z)$.  The corresponding distances are calculated using Eq.~\ref{eq:comovingr}. The number of flat galaxies, N$_{flat}$, should be reasonably high since they are later used as seed galaxies for generating more galaxies to mimic the gravitationally bound galaxy clusters, ``clumps''.

\begin{figure}[h!]
    \centering
    \ifdefined\colored
        \includegraphics[width=.55\textwidth]{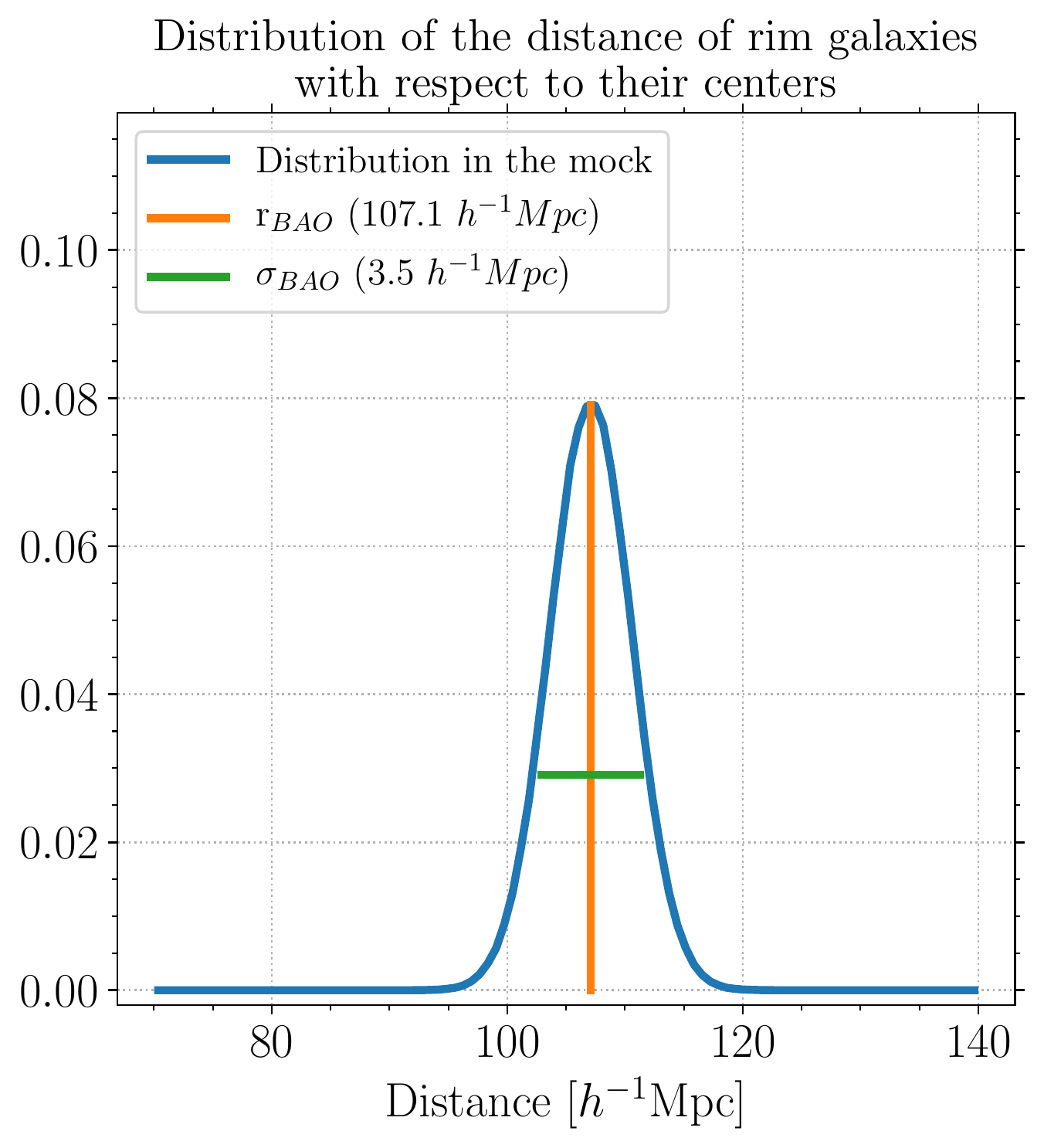}
    \else
        \includegraphics[width=.55\textwidth]{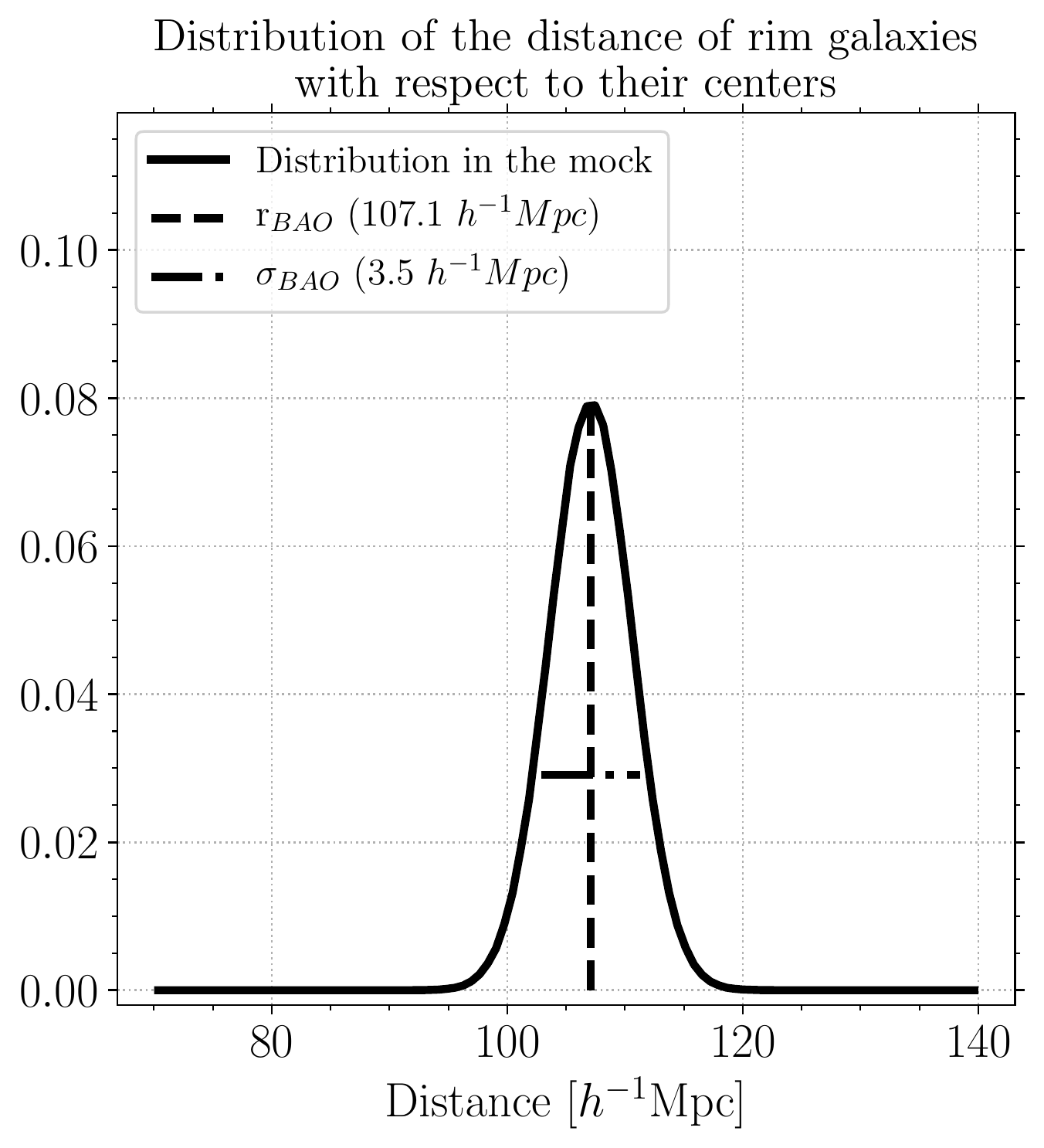}
    \fi
    \caption{The distribution of distance of the rim galaxy to their corresponding center galaxies. The dashed vertical and the dotted horizontal lines show the input values for a BAO distance, $r_{BAO}$, and a smearing of the distance, $\sigma_{BAO}$.}
    \label{fig:BAO_distance_distribution}
\end{figure}

\paragraph{Center and Rim Galaxies:}
To inject BAO signal into the mocks, we randomly generate N$_{center}$ ``centers'' for BAO ``bubbles'' according to the uniform angular distribution and $P(z)$. Then, for each BAO center, we add $N_{rim}$ ``rim'' galaxies displaced from the center by a distance that is normal distributed with a mean of $r_{BAO}$ and width of $\sigma_{BAO}$ (See Figure \ref{fig:BAO_distance_distribution}). Rim  galaxies are uniformly distributed over the solid angle around the associated BAO center.  Some of the rim galaxies can be discarded to correct for the angular mask and redshift distribution. 

\paragraph{Clump Galaxies:}
``Clumping'' is introduced around $NR_{clump}$ galaxies randomly drawn from the newly-created mock catalog to mimic galaxy clusters. The number of clump galaxies to be placed around the seed galaxy is pulled from a Poisson distribution with BAO centers having a different mean number of clumping partners because of the potential access of dark matter around the BAO centers. The distance $r$ of the clump galaxies to their respective seed galaxy follows the distribution \ \cite{Peebles:2001cy}:
\begin{equation} \label{eq:clump}
    f(r) = r_{scale} P_{pareto}\left(u; \gamma\right) %%A\times(r_0/r)^{\gamma},
\end{equation}
where $u$ is random number in [0, 1],  $r_{scale}$ and $\gamma$ are input parameters. The P$_{pareto}$ is the Pareto distribution given by:

\begin{equation}
    P_{pareto}(u; \gamma) \propto \left( u ^ {-\gamma-1} \right)
\end{equation}

\begin{figure}
    \centering
    \ifdefined\colored
        \includegraphics[width=.55\textwidth]{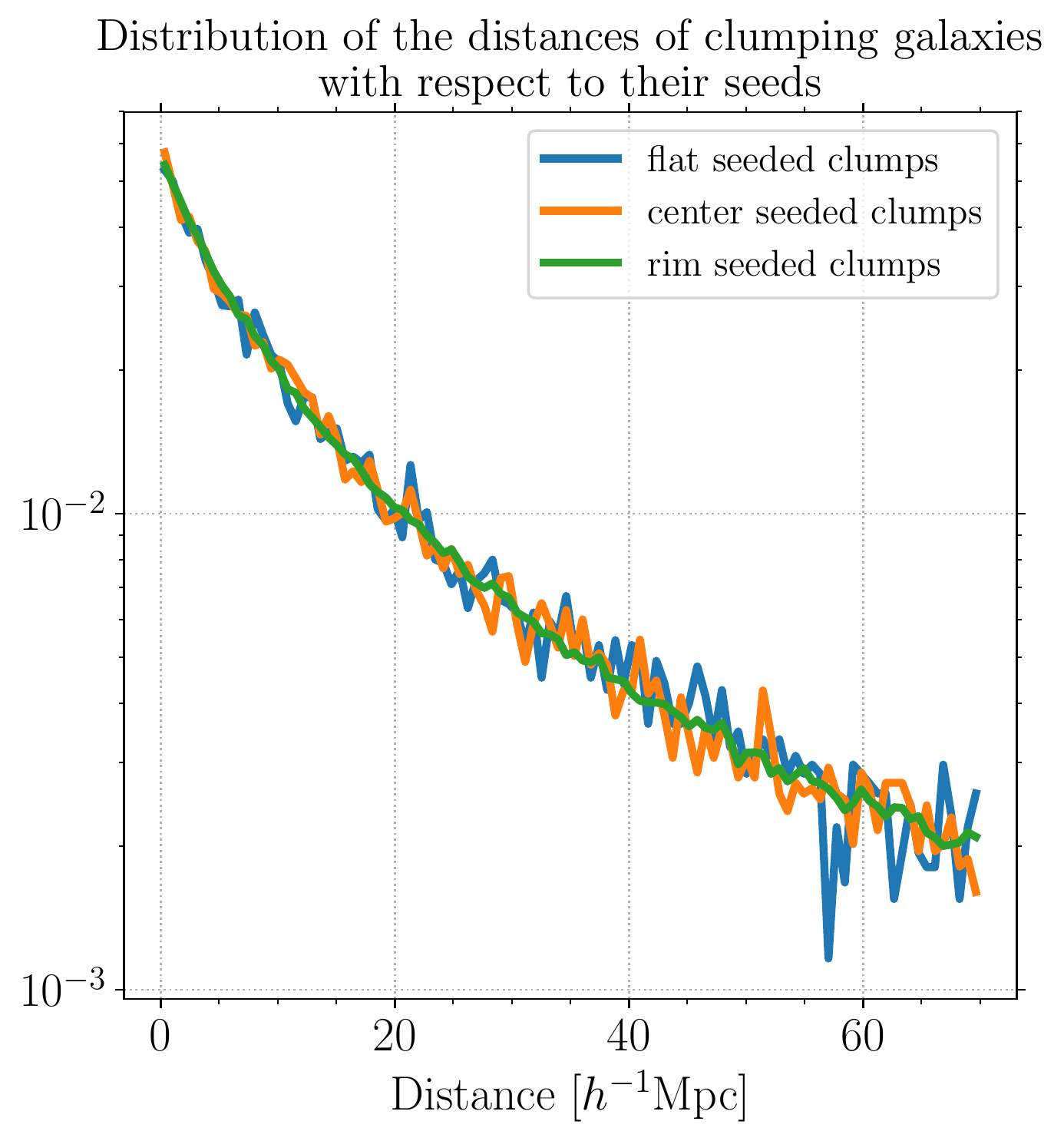}
    \else
        \includegraphics[width=.55\textwidth]{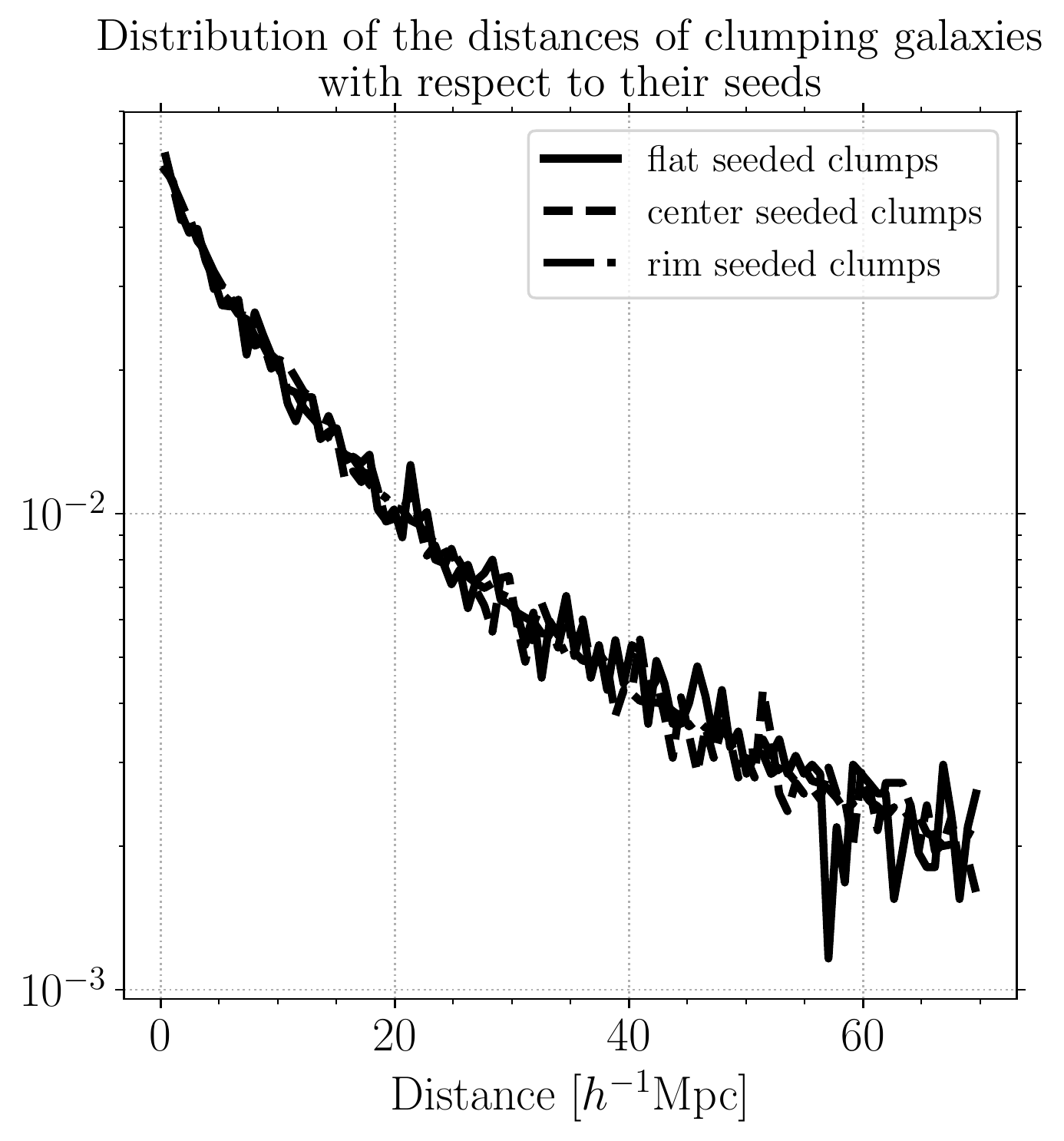}
    \fi
    \caption{The distribution of the distance of clumped galaxies to their seeds for different categories of seeds. It is described by Equation~\ref{eq:clump}. The variations at the tail of the distributions are due to statistics.}
    \label{fig:clump_distribution}
\end{figure}

Depending on the value of $r_{scale}$, the model is capable of generating galaxy clusters and superclusters. Examples of the  distributions of different categories of clumped galaxies over their distance to the seed galaxy are shown in Figure \ref{fig:clump_distribution}.

\begin{figure}[h!]
\begin{center}
\ifdefined\colored
    \includegraphics[width=.65\linewidth]{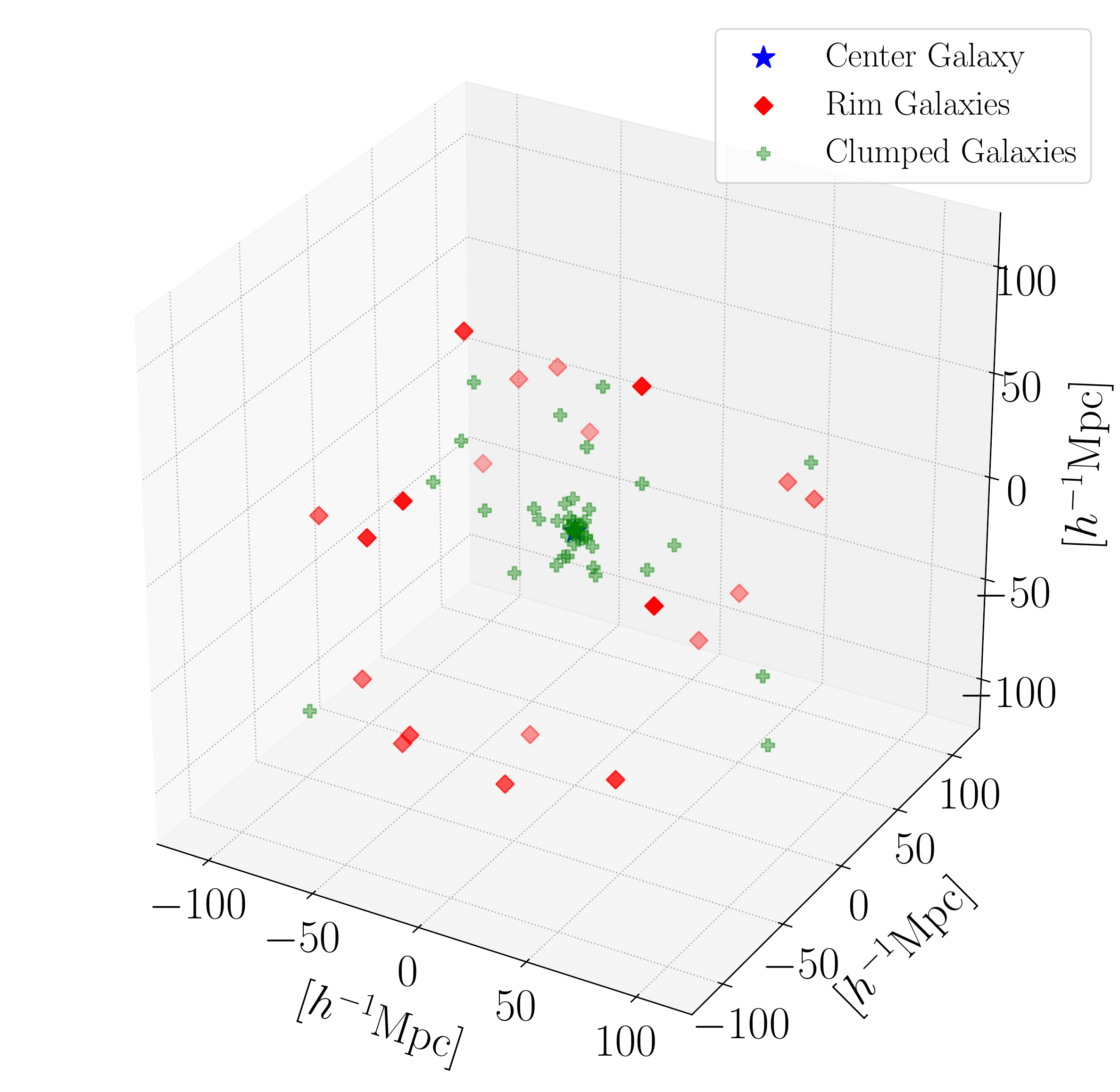}
\else
    \includegraphics[width=.65\linewidth]{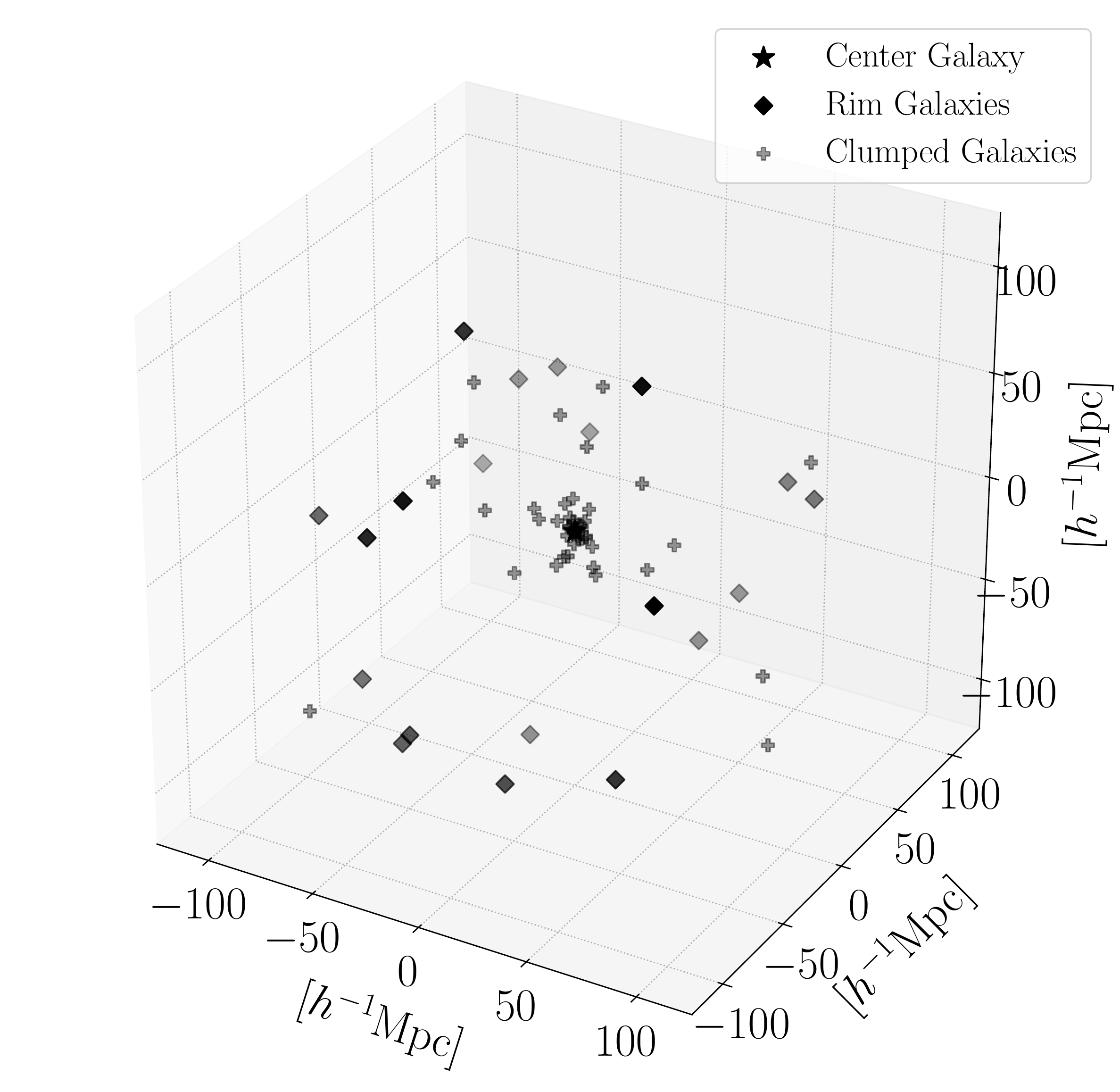}
\fi
\end{center}
%\vspace{-3.7cm}
\ifdefined\colored
    \caption{A small sample of galaxies generated with the model. The blue point at the center is a center galaxy in the mock catalog. The center galaxy is surrounded by rim galaxies, shown in red, following a normal distribution with a mean of 107.1 $h^{-1}$Mpc and sigma of 3.5 $h^{-1}$Mpc. Clump galaxies following the distribution described by Equation~\ref{eq:clump} are shown in green.}
\else
    \caption{A small sample of galaxies generated with the model. The point (marked with `star') at the center is a center galaxy in the mock catalog. The center galaxy is surrounded by rim galaxies, shown with diamond symbols, following a Gaussian distribution with a mean of 107.1 $h^{-1}$Mpc and smearing of 3.5 $h^{-1}$Mpc. Clump galaxies following the power law are shown with light gray diamonds.}
\fi
\label{fig:BAO_cartoon}
\end{figure}

A cartoon in Fig. \ref{fig:BAO_cartoon} shows a single center with the associated rim and clumped galaxies. Not all the center galaxies and flat galaxies are selected to have clumps but all center galaxies have the rim around them.

% The co-moving coordinates $\vec{r}$ of any galaxy can be described by its right
% ascension $\alpha$, declination $\delta$, and co-moving radial distance $r$.
% Since random catalogs are typically generated by uniformly populating the
% fiducial volume of the survey, the galaxy distribution over $z$ and thus, over
% $r$, is factorizable from the angular distribution. In other words, any angular
% region of the sky has the same distribution of galaxies in $r$ (see
% \cite{Ross:2012qm} ). This means that the expected count of random galaxies,
% $R(\vec{r})$, can be factorized into the product of the expected count
% $R_\text{ang}(\alpha,\delta)$ at a given angular position and a radial
% probability density function (PDF) $P_r(r)$:
% %
% \begin{linenomath}
% \begin{equation} \label{eq:factorized}
%   R(\vec{r}) =
%   R_\text{ang}(\alpha, \delta)
%   P_r(r).
% \end{equation}
% \end{linenomath}

\subsection{\boldmath Modeling Redshift-Space Distortions}
\label{sec:RSD}

In spectroscopic surveys, the radial distances to the galaxies are anisotropically distorted due their peculiar motions. In order to account for this distortion, we change the shape of the BAO bubble by moving the galaxies along the LOS closer or farther to the respective center galaxy, effectively changing the spherically shaped BAO bubble into a spheroid. We keep our model as general as possible so that the user has an option to distort the BAO bubble in directions other than along LOS. 

\begin{figure}
    \centering
    \ifdefined\colored
        \includegraphics[width=.48\textwidth]{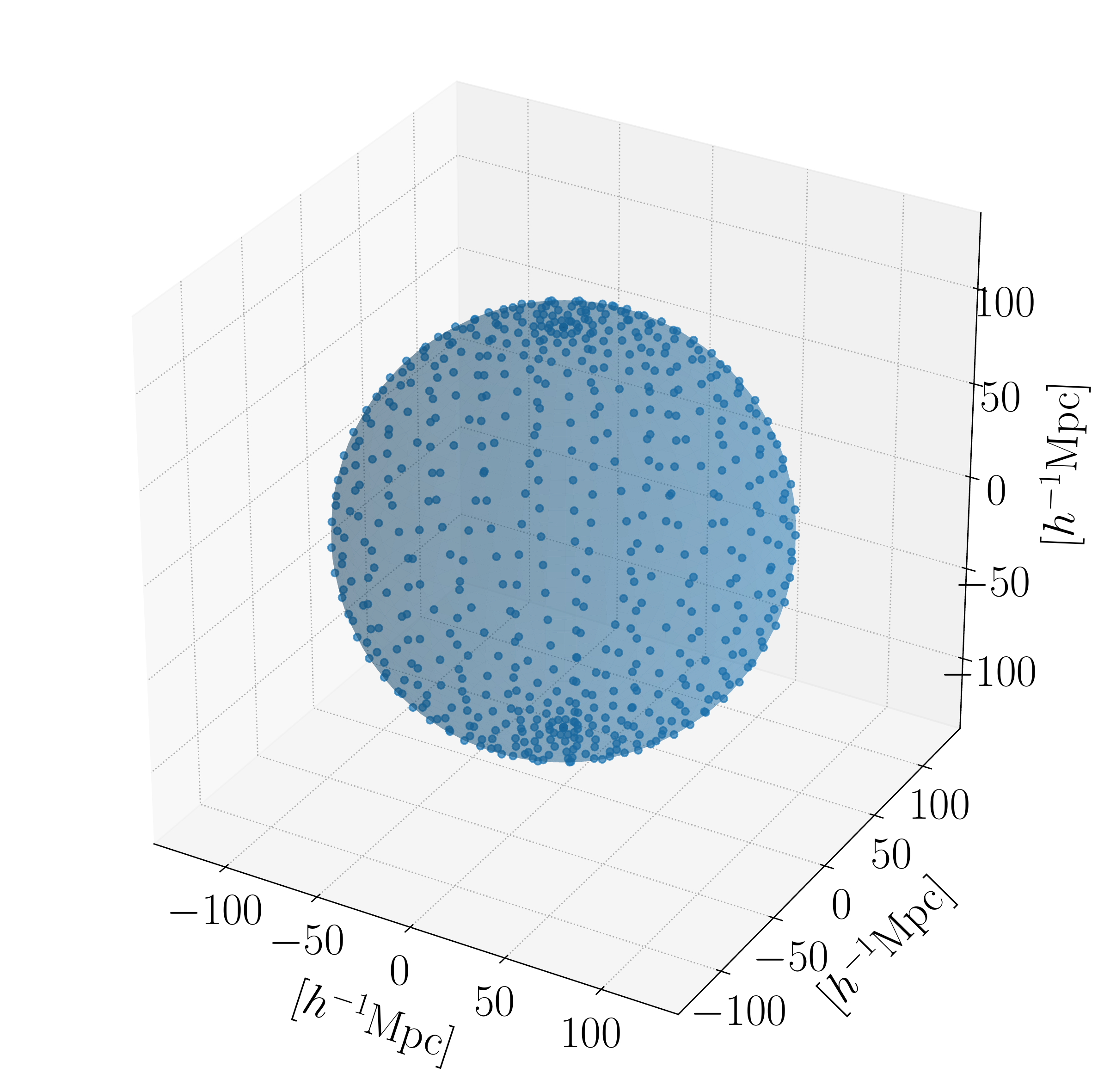}
        \includegraphics[width=.48\textwidth]{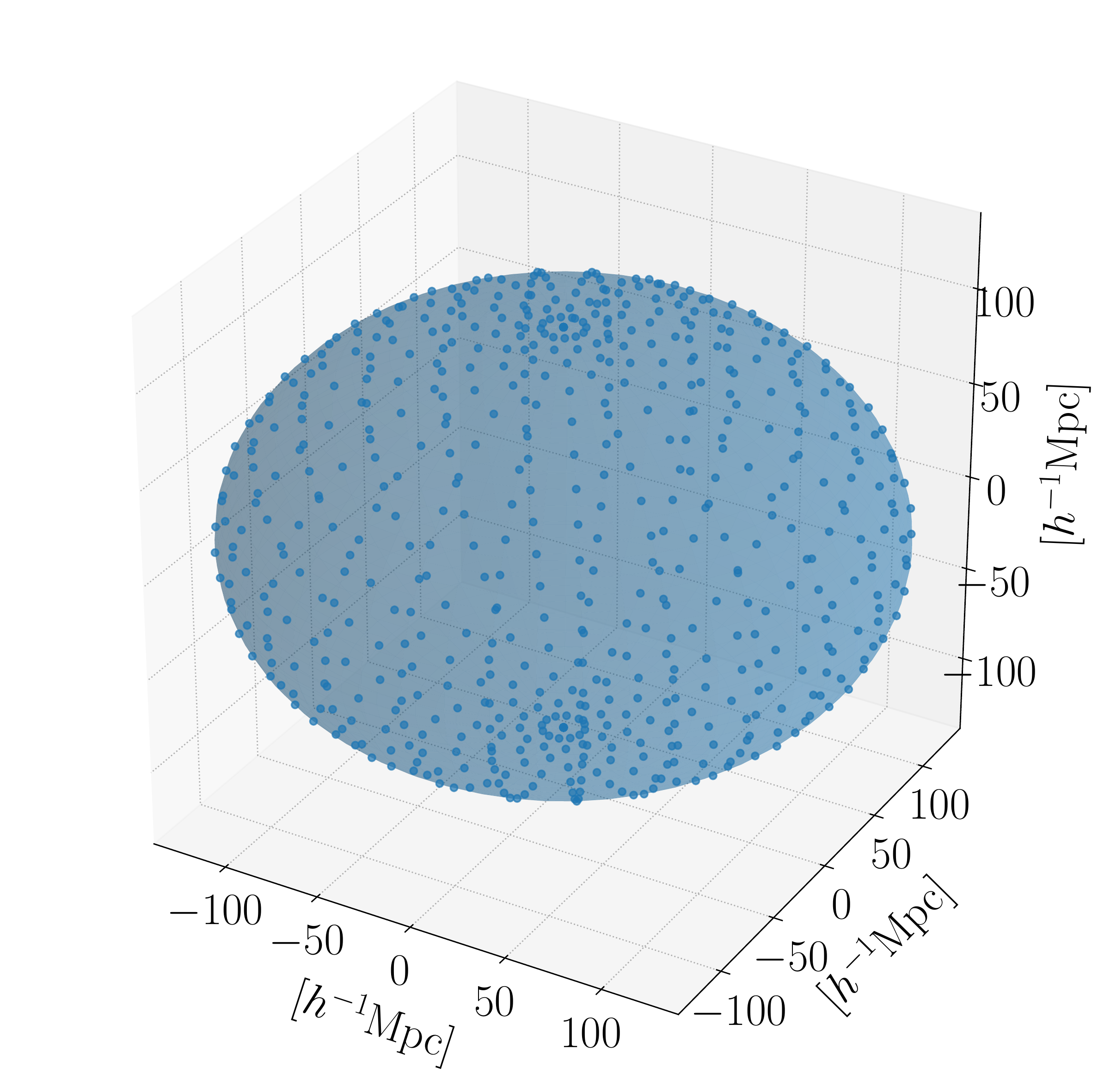}
    \else
        \includegraphics[width=.48\textwidth]{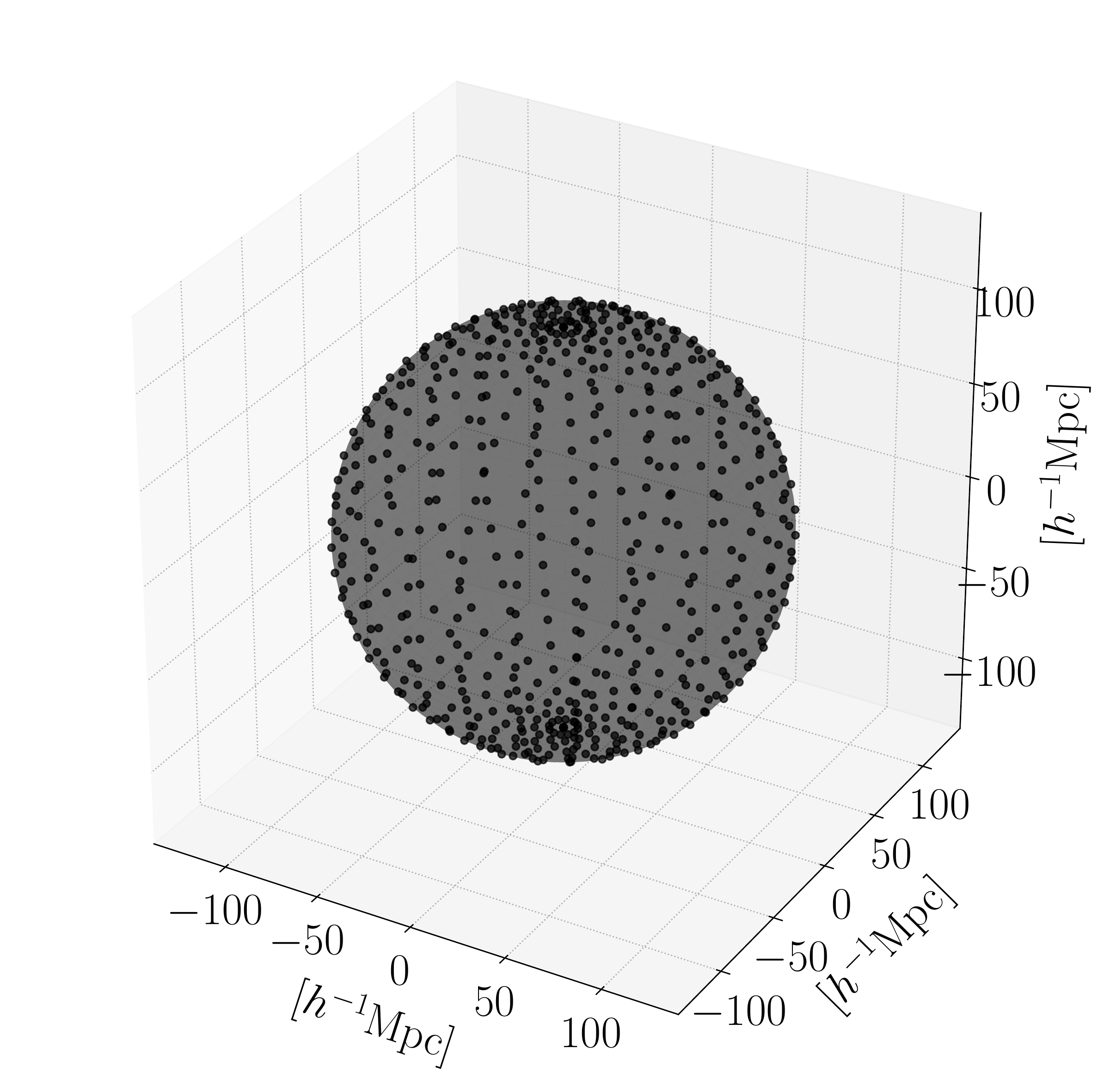}
        \includegraphics[width=.48\textwidth]{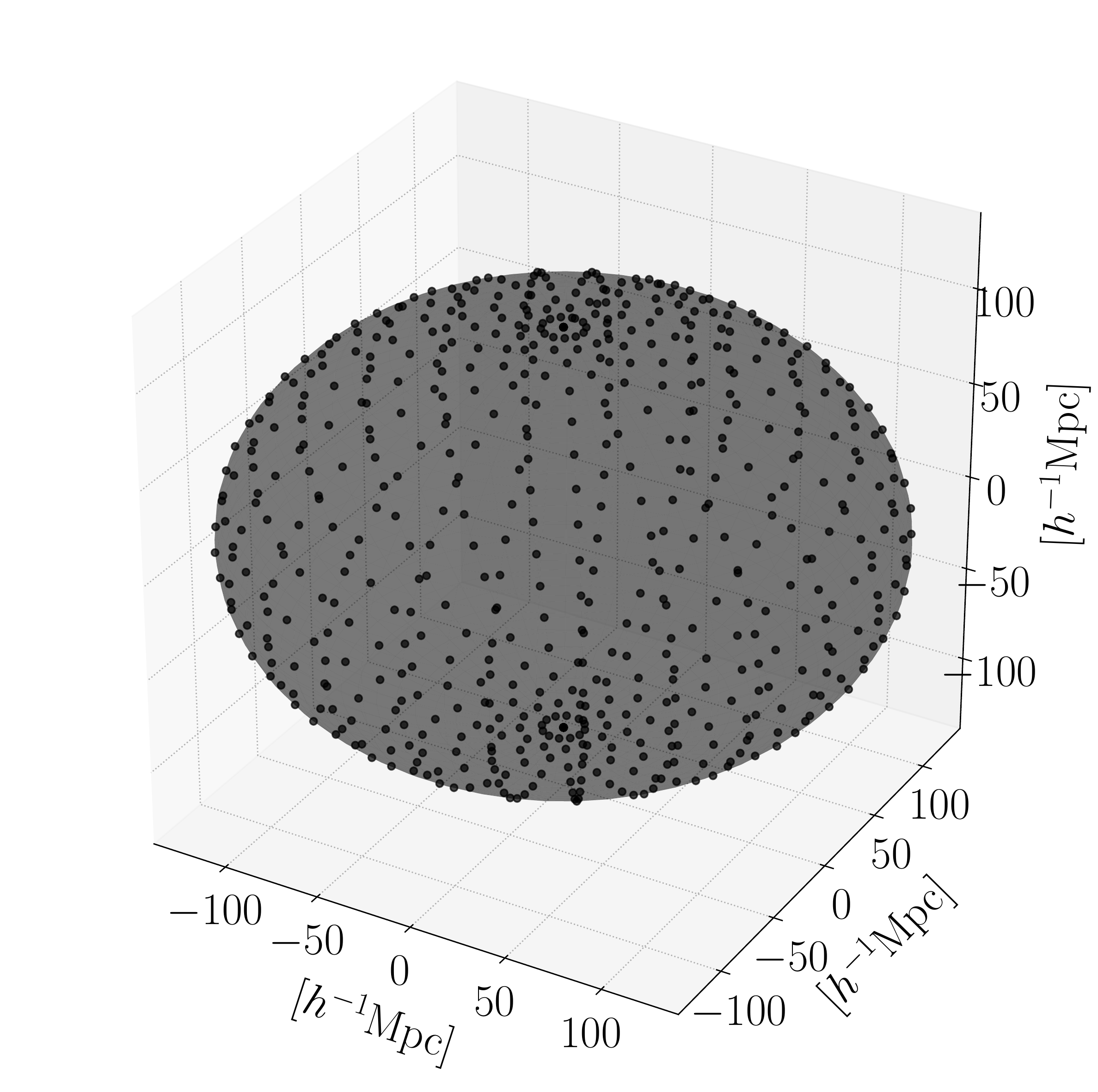}
    \fi
    \caption{Illustration of the BAO bubbles modeling. \textit{(left)} A spherical BAO bubble in the absence of RSD. \textit{(right)} A spheroid BAO bubble distorted by the RSD.}
    \label{fig:example_BAO_bubbles}
\end{figure}

We first generate the BAO bubbles around their respective center galaxies using a local cartesian coordinate system (centered on the bubble center), then we transform the bubbles into global coordinates. In the case of the anisotropic BAO bubbles, we also  rotate the BAO bubble to ensure the proper LOS orientation.  
%Doing so guarantees, for each BAO bubble, LOS is actually on the line-of-sight rather than being in the wrong direction because of the method we generate the BAO bubble. 
Examples of the isotropic and anisotropic BAO bubbles are shown in Figure \ref{fig:example_BAO_bubbles}. 

\subsection{\boldmath Acceptance Adjustment}
\label{sec:acceptance}

It is important to guarantee that the newly produced data follow the user-defined redshift distribution, $P(z)$. Suppose the newly added galaxies follow the distribution $P_{\text{mock}}(z)$. Then, some galaxies need to be rejected to reproduce the original distribution $P(z)$. This procedure is referred to as the acceptance adjustment procedure.

\begin{figure}[h!]
    \centering
    \includegraphics[width=.6\textwidth]{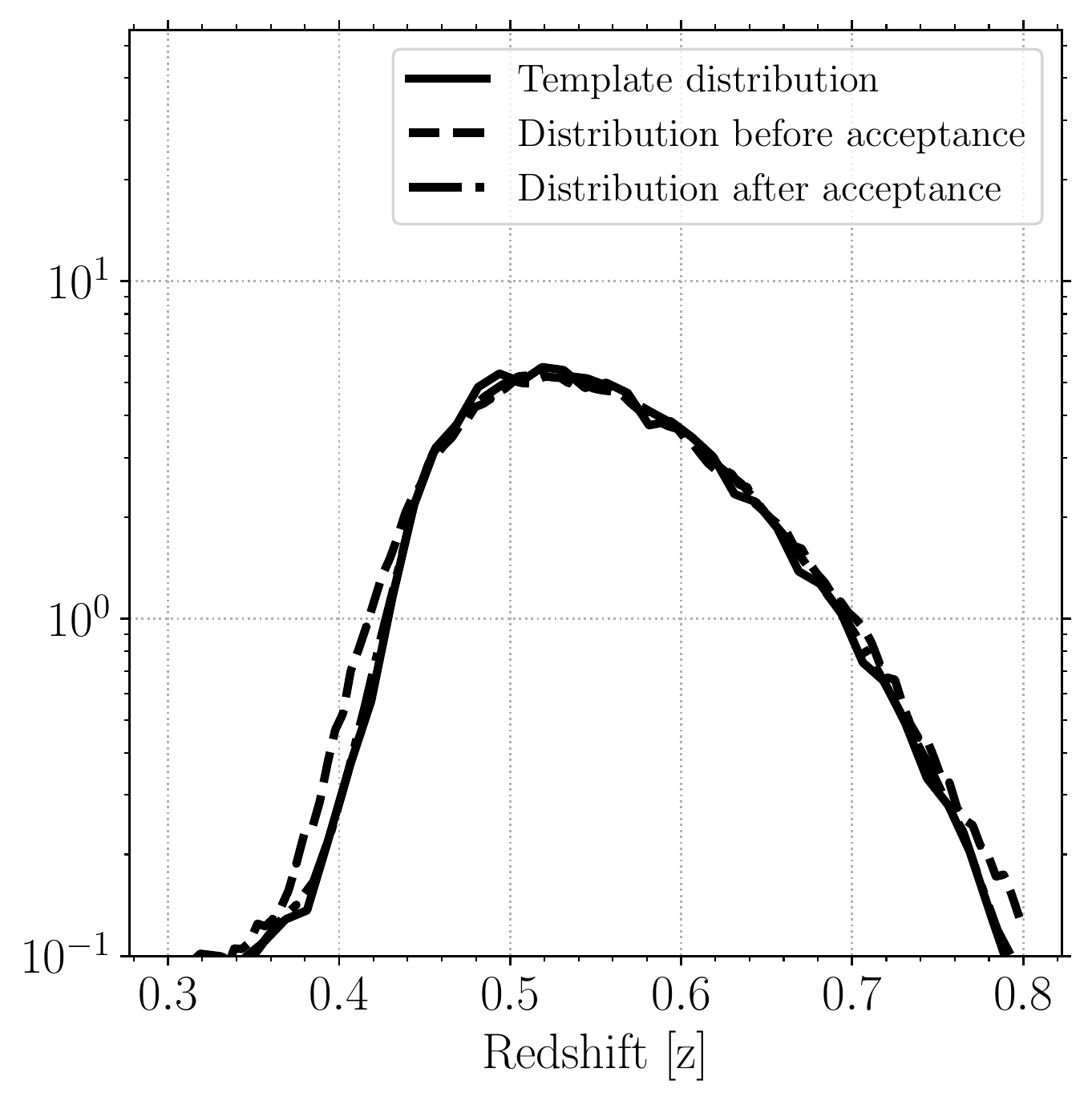}
    \caption{ Redshift distribution of the galaxies generated by the algorithm before and after the acceptance adjustment. The input template distribution is recovered after the acceptance adjustment procedure is applied.}
    \label{fig:acceptance_stat}
\end{figure}

We introduce a positively defined function $\epsilon(z)=P(z)/P_{\text{mock}}(z)$ with $\epsilon_{max}$ being its maximum value in the region of $z$ under investigation. For every newly generated galaxy with a redshift $z$, we throw a random number $x$ uniformly distributed between zero and $\epsilon_{max}$. If this number is greater than $\epsilon(z)$ the galaxy gets removed from the sample, otherwise it is kept. The resultant distribution of remaining galaxies is $P(z)$. An example of the result of this procedure is shown in Figure \ref{fig:acceptance_stat}. The efficiency of the acceptance adjustment procedure  is typically $\sim$97\%.

\section{\boldmath Performance of the algorithm}
\label{sec:performance}

\begin{figure}
    \centering
    \includegraphics[width=.6\textwidth]{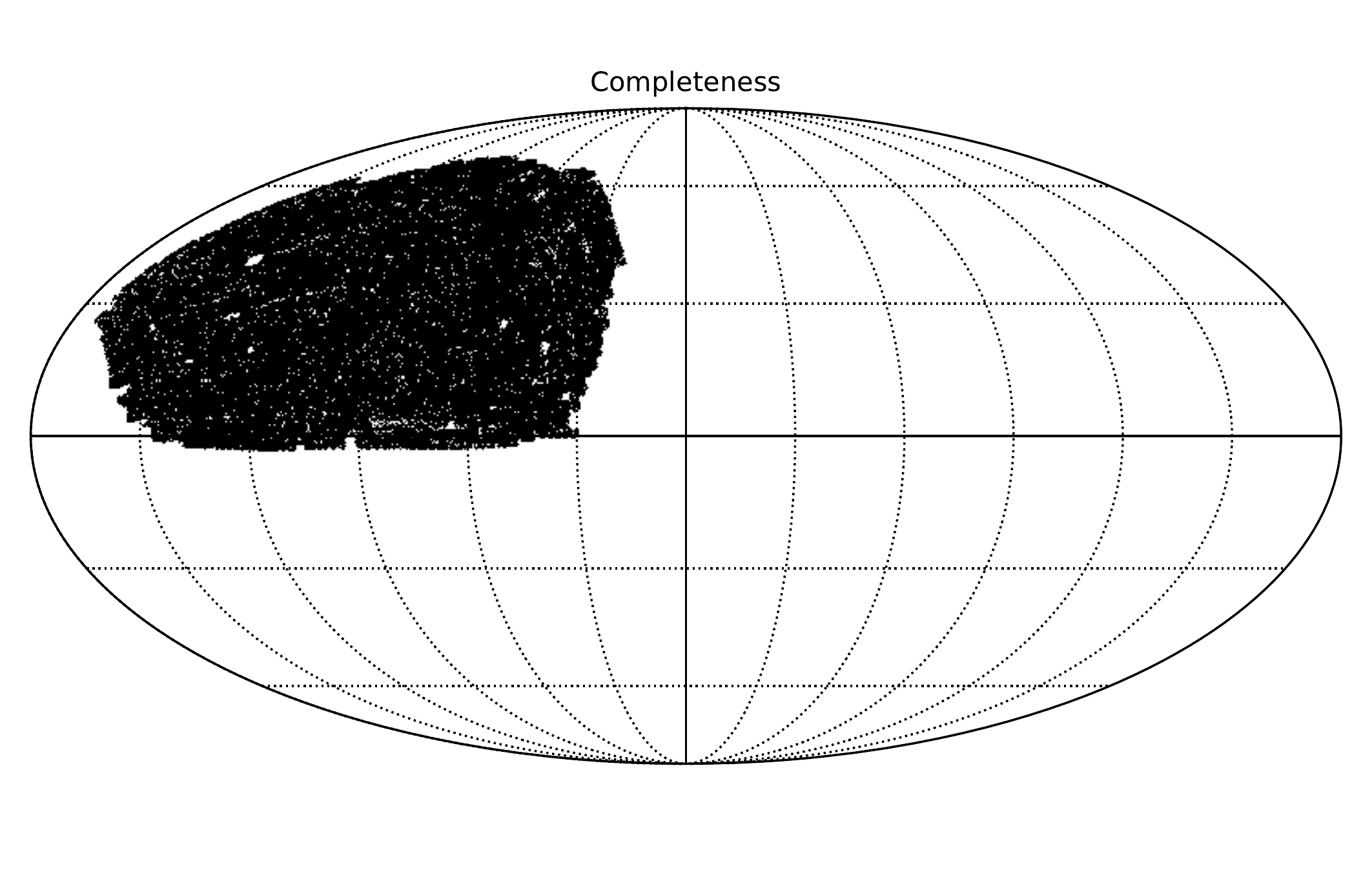}
    \caption{The angular mask (marked by blue region) used for generating the mock catalogs with parameters from in Table \ref{tbl:parameters}.}
    \label{fig:angmask}
\end{figure}

We generate several mock catalogs with parameters listed in Table~\ref{tab:perf} to demonstrate the performance of the method.  These parameters are not intended to model the distributions in any specific catalog. They are rather chosen to highlight certain features of different type of galaxies in the performance measures. The user-defined redshift distribution and the angular mask are shown in Figures \ref{fig:acceptance_stat} and \ref{fig:angmask}. A random catalog is also generated with the same parameters as \mockone but with $\sim$5 times the number of galaxies. As performance measures, we use two-point correlation function (2pcf), and the power spectrum (PS).

\begin{table}[htbp]
\caption{Parameters of mock data catalogs.\label{tab:perf}}
\begin{center}
\begin{tabular}{lcccc}
Parameter               & Mock 1 [flat] & Mock 2 [clumps] & Mock 3 [rims] & Mock 4 [complete]\\
\hline\hline 
$N_{center}$            & 0     & 0    & 30k  & 30k  \\ 
$n_{rim}$               & 0     & 0    & 10   & 10   \\ 
$N_{flat}$              & 750k  & 550k & 450k & 250k \\ 
$NR_{clump}$            & 0     & 20k  & 0    & 20k \\ 
$n_{clump}$             & 0     & 10   & 0    & 10   \\ 
$n_{centerclump}$       & 0     & 10   & 0    & 10   \\
\hline
$\gamma$                & \multicolumn{4}{c}{1.8}    \\ 
$r_{min} [h^{-1}Mpc]$   & \multicolumn{4}{c}{2.1}    \\ 
$r_{scale} [h^{-1}Mpc]$ & \multicolumn{4}{c}{5.6}    \\ 
\hline
$R_{BAO} [h^{-1}$Mpc]   & \multicolumn{4}{c}{107.1}  \\ 
$\sigma R_{BAO}[h^{-1}Mpc]$& \multicolumn{4}{c}{3.5} \\ 
\hline
$h$                     & \multicolumn{4}{c}{0.7} \\
$\Omega_{M}$            & \multicolumn{4}{c}{0.31} \\
$\Omega_\Lambda$        & \multicolumn{4}{c}{0.69}  \\ 
\hline
\end{tabular}
\label{tbl:parameters}
\end{center}
\end{table}

\subsection{Two-Point Correlation Function}

We use two-point correlation function (2pcf) $\xi(s)$ introduced by Peebles to quantify the large scale structure produced by the discussed model. Given a spectroscopic survey containing the 3D coordinates of each galaxy there exist several possible ways to estimate $\xi(s)$. The most popular estimator is due to Landy-Szalay\cite{Landy:1993yu} (see also \cite{Hamilton:1993fp}), which is constructed by combining pairs of galaxies from a catalog of measured objects $D$ (``data'') and from a randomly generated catalog $R$ with galaxies distributed uniformly over the fiducial volume of the survey but using the same selection function as the data. The Landy-Szalay estimator is
\begin{equation}\label{eq:2pcf}
  \hat{\xi}(s) = \frac{DD(s) - 2DR(s) + RR(s)}{RR(s)},
\end{equation}
where $DD$, $RR$, and $DR$ are the normalized distributions of the pairwise combinations of galaxies from the data and random catalogs (plus cross terms) at given distances $s$ from each other. The distances $s$ is calculated from the component parallel to the LOS, $\sigma$, and the transverse component, $\pi$ as explained in \cite{Demina:2016gxl}:

\begin{eqnarray}
\sigma_{12} & = & \left( t_1 + t_2 \right) \sin{\frac{\theta_{12}}{2}} \\
\pi_{12} & = & \mid r_1 - r_2 \mid \cos{\frac{\theta_{12}}{2}}
\end{eqnarray}
where $\theta_{12}$ is the angular separation between the two galaxies. $r$ is the radial distance along LOS evaluated according to Equation~\ref{eq:comovingr}) and  $t$ is the transverse to LOS distance given (under the assumption of small $\Omega_k$) by: 
\begin{eqnarray}
%r(z) & = & \frac{c}{H_0} I(z) \\
t(z) & = & r(z)\left(1+\frac{\Omega_k}{6}(I(z))^2\right).
\end{eqnarray}

We calculated the 2pcf using a python package called \texttt{KITCAT}\footnote{https://github.com/DESI-UR/KITCAT}\cite{nguyen_2019_2640917} developed based on~\cite{Demina:2016gxl}. We use 4 Mpc for the binning in $s$ and we let \texttt{KITCAT} calculate the other configuration parameters such as angular binnings following the prescription in~\cite{Demina:2016gxl} automatically.

\begin{figure}[h!]
\centering
\includegraphics[width=\linewidth]{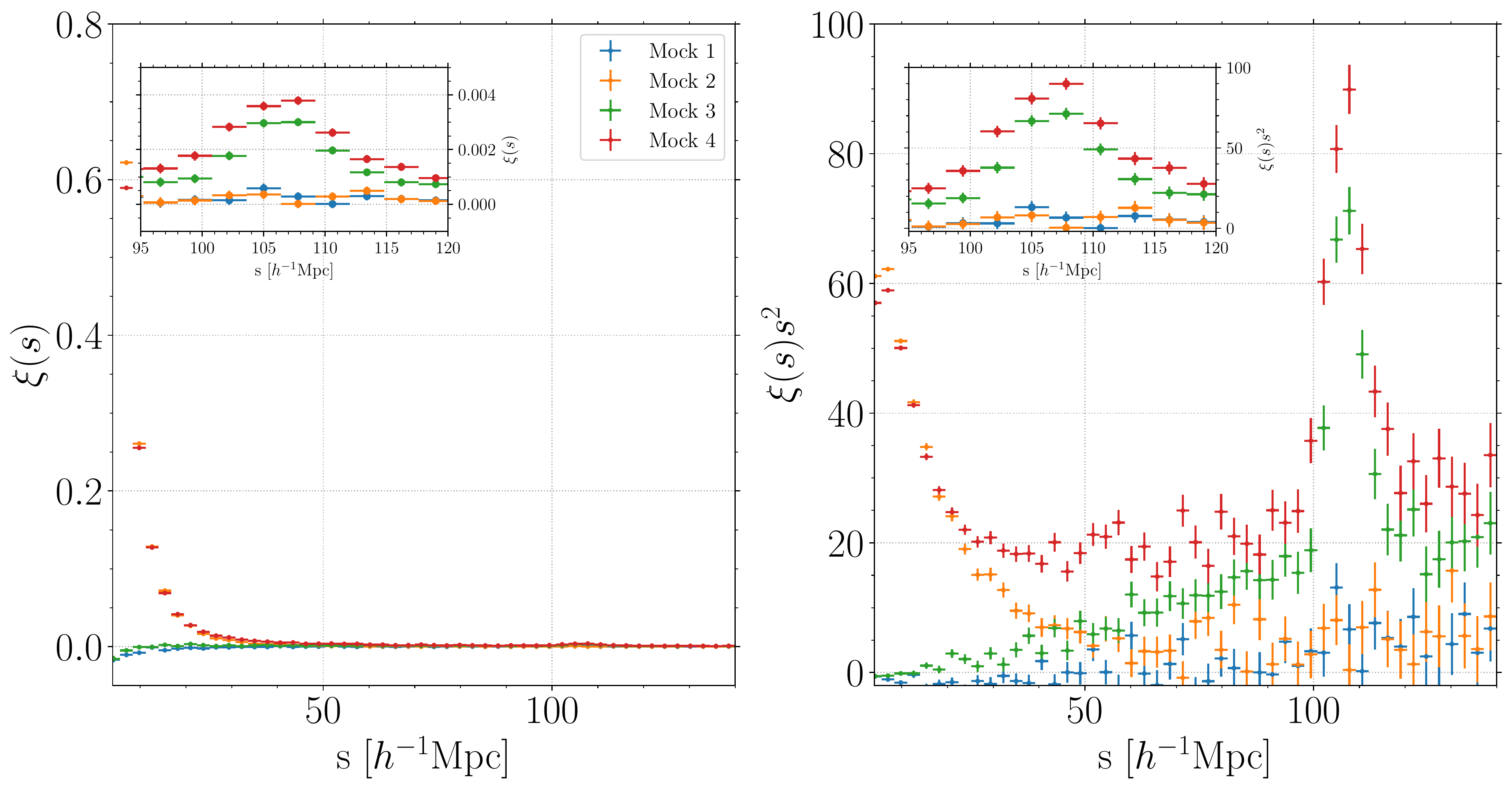}
\caption{\label{fig:allresults} 2pcf(left) and $2pcf s^2$(right) for the generated sample mocks described in Table \ref{tbl:parameters}. Inserts show zoomed up region around $R_{BAO}$}
\end{figure}

In Figure~\ref{fig:allresults}, we show the one-dimensional 2pcf result for \mockone, \mocktwo, \mockthree and \mockfour. \mockone shows no features whereas a definitive galaxy clustering correlation is shown for \mocktwo. \mockthree, that has only uniformly distributed galaxies and added BAO bubbles, shows the characteristic BAO peak at the user-defined distance. As expected \mockfour demonstrates all the features, including gravitational clumping feature at small distances and the BAO peak at user-defined distance. 

\begin{figure}[h!]
    \centering
    \includegraphics[width=.48\linewidth]{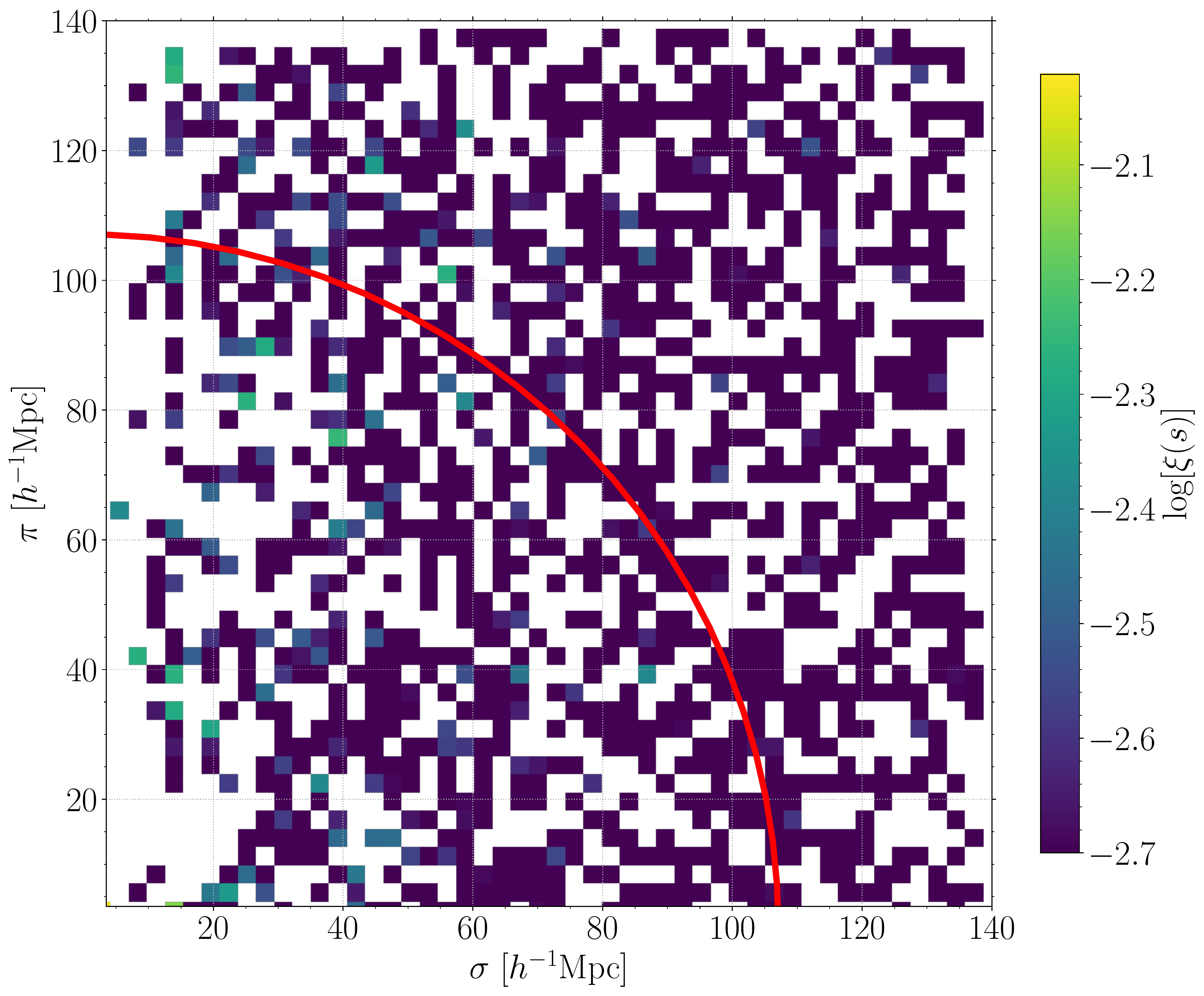}
    \includegraphics[width=.48\linewidth]{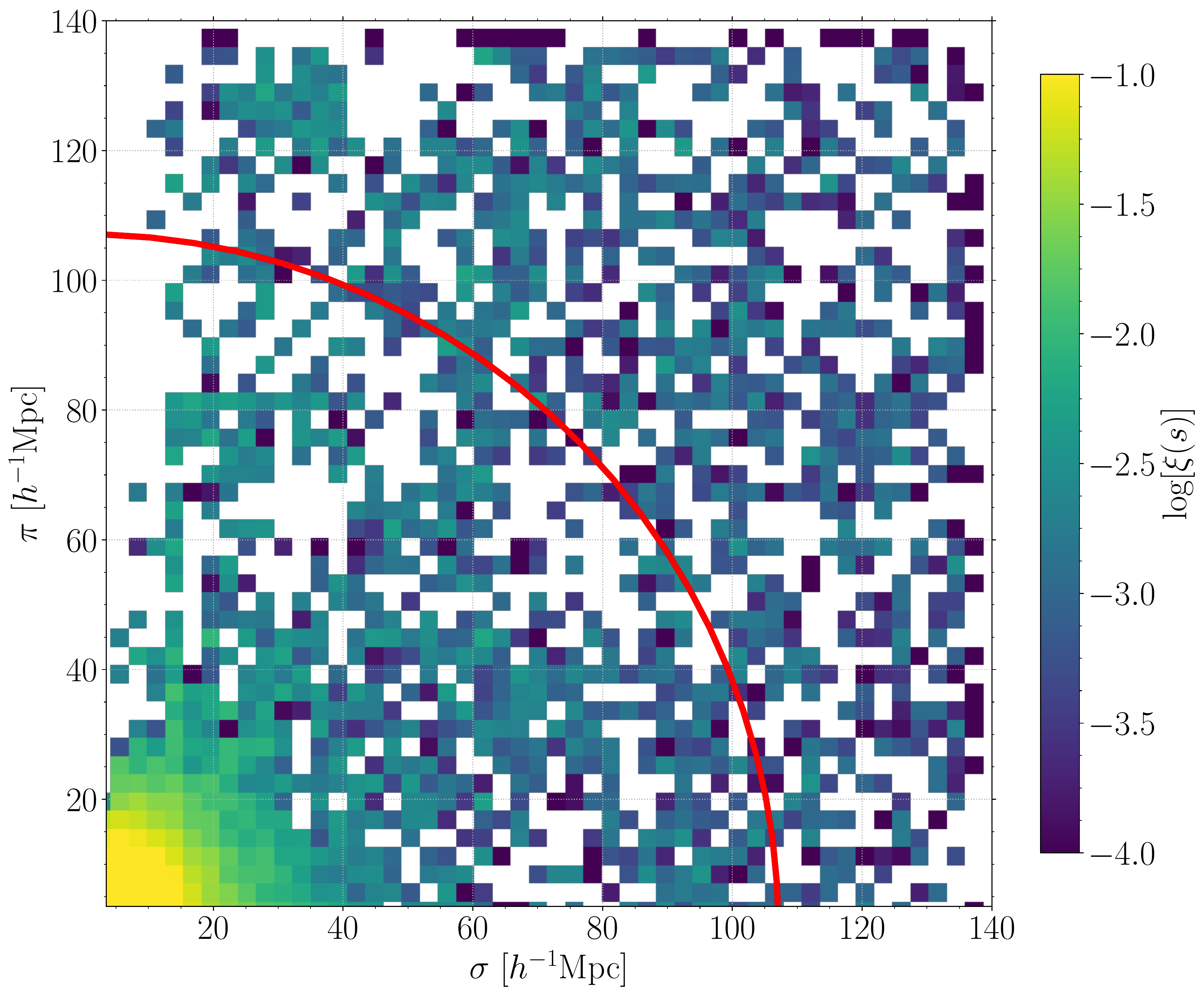}\\
    \includegraphics[width=.48\linewidth]{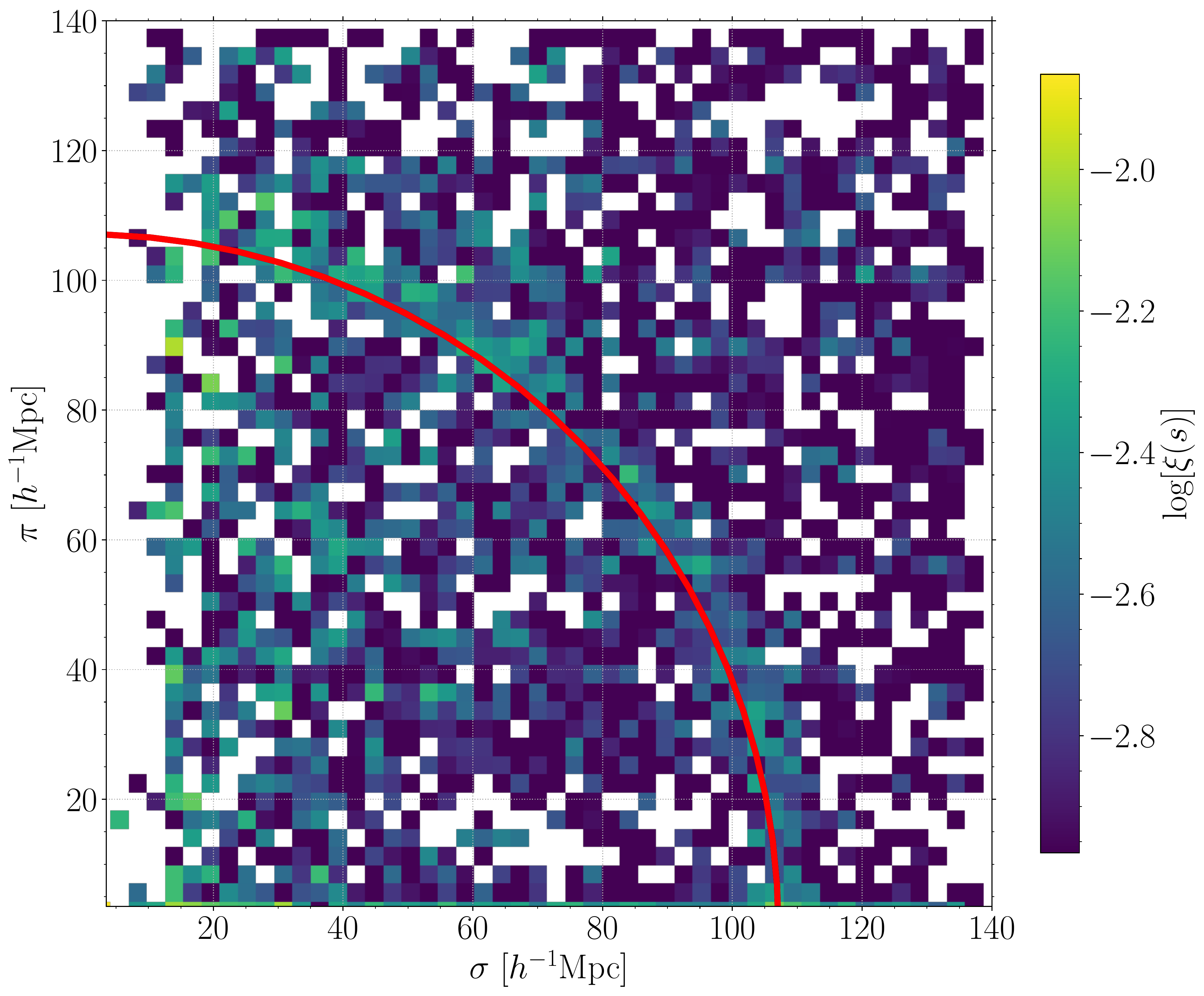}
    \includegraphics[width=.48\linewidth]{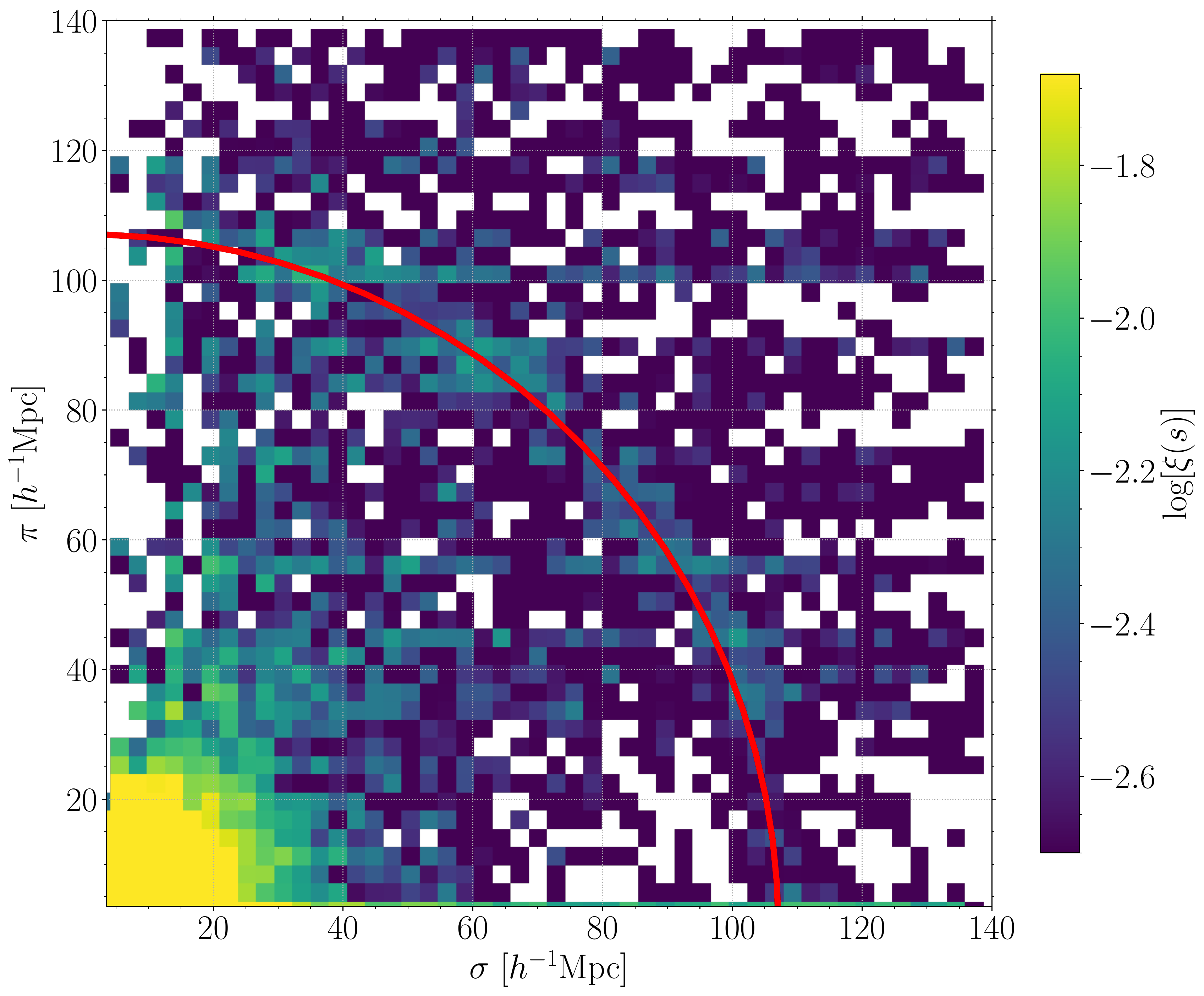}
    \caption{\label{fig:2d_2PCF} 2D 2pcf for \mockone (top-left), \mocktwo (top-right), \mockthree (bottom-left) and \mockfour (bottom-right)}
\end{figure}

\begin{figure}[h!]
    \centering
    \includegraphics[width=.48\linewidth]{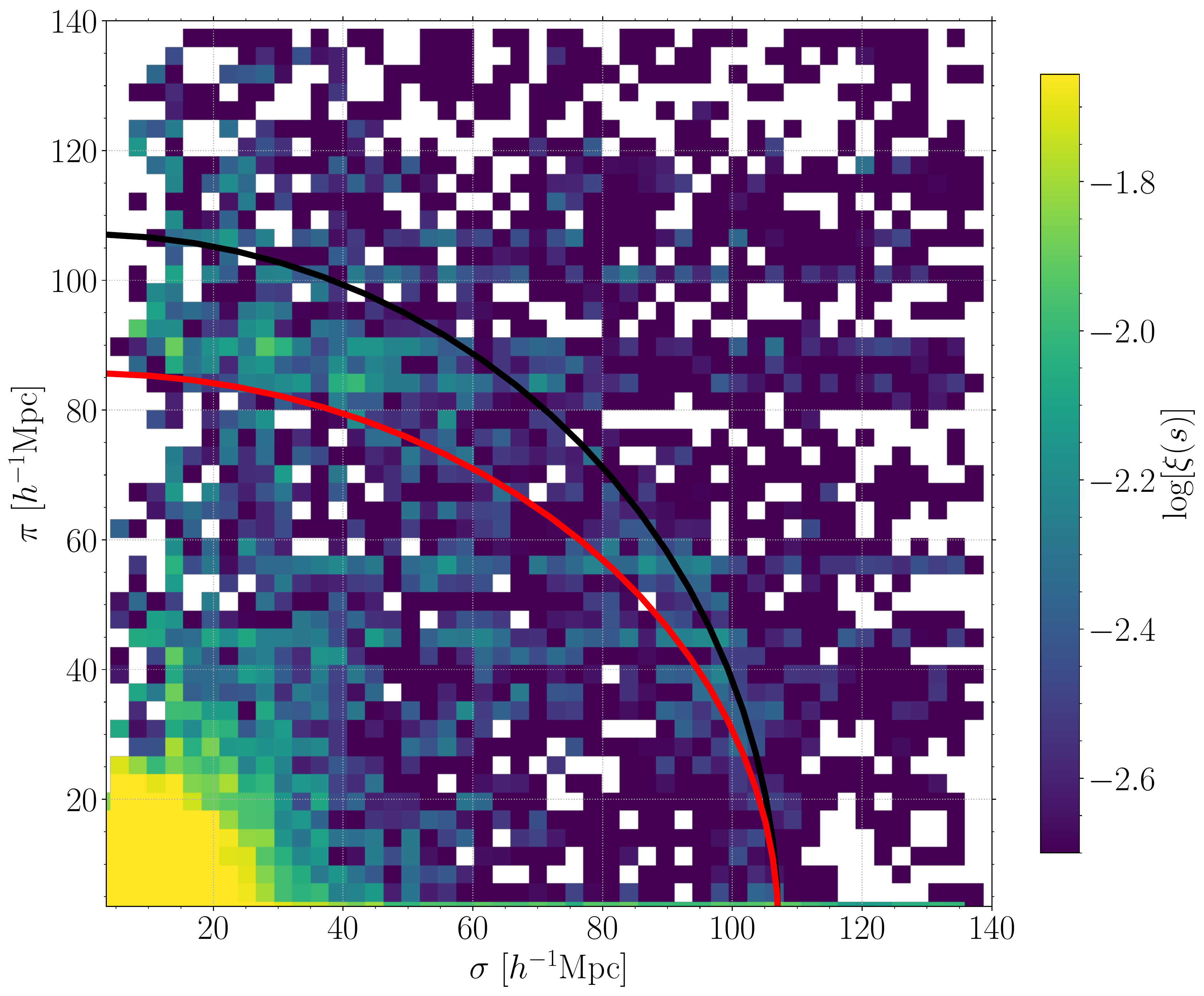}
    \includegraphics[width=.48\linewidth]{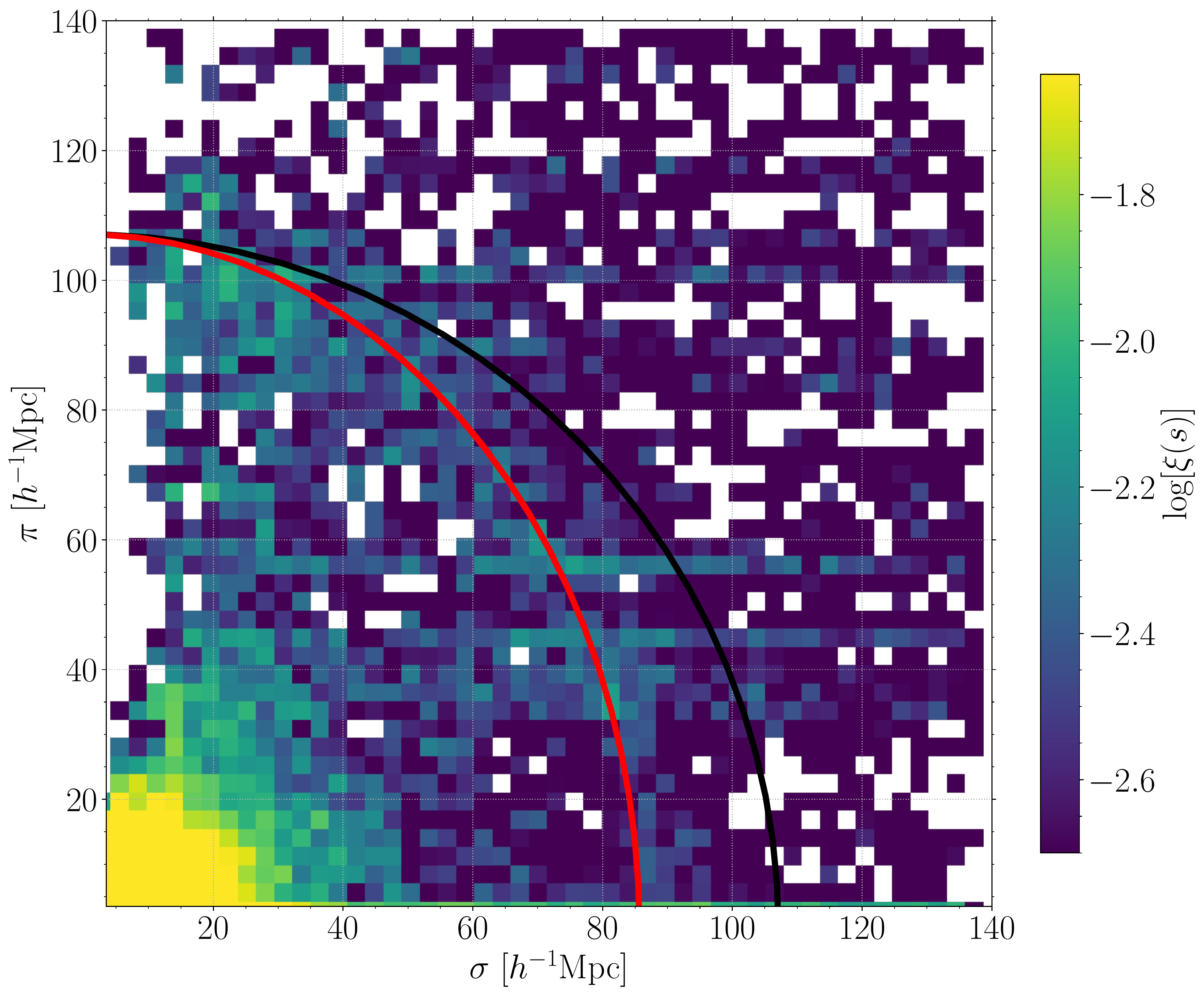}
    \caption{\label{fig:2danisotropy} 2D 2pcf for the anisotropic mocks in which the BAO bubbles are stretched along the LOS(left) and perpendicular to the LOS(right). Red quarter circle shows the user-defined BAO distance with the RSD and the blue line shows the same BAO bubble if there were no RSD.}
\end{figure}

Similar features are present in the 2D 2pcf  of these mocks (Figure~\ref{fig:2d_2PCF}). The correlation values are plotted in logarithmic scale to enhance visual comparison. \mockthree demonstrates the imprinted isotropic BAO bubble with no redshift distortions (bottom-right panel of Figure \ref{fig:2d_2PCF}). Note that the red lines in the figures show the expected BAO feature and the grey line shows the BAO feature in the absence of RSD for the given conditions. For testing of the anisotropy, we distort the BAO spheres in the mock generation as explained in Section \ref{sec:RSD}. We generate two new mock catalogs based on the parameters used for \mockfour: a mock catalog with BAO bubbles distorted along the LOS and a mock catalog with BAO bubbles distorted perpendicular to the LOS (Figure \ref{fig:2danisotropy}). The results meet our expectations and follow the expected BAO bubble aspects as shown in the figures.

% \subsection{Three-Point Correlation Function}

% Three point correlation function (3PCF) provides further information for understanding the cosmological parameters. Thus, we use 3PCF as another performance measure for our model. The generalized form of the N-PCF estimator (Equation \ref{eq:Npcf}) is given by \cite{Szapudi_1998}:

% \begin{equation}\label{eq:Npcf}
%   \hat{\xi} = \frac{ (D_1 - R_1)(D_2 - R_2)...(D_N - R_N) }{R_1R_2...R_N},
% \end{equation}
% Reducing Equation \ref{eq:Npcf} to three points, the estimator is written as :

% \begin{eqnarray}\label{eq:3pcf}
%   \hat{\xi} & = & \frac{ (D_1 - R_1)(D_2 - R_2)(D_3 - R_3) }{R_1R_2R_3} = \frac{ NNN }{  RRR }
% \end{eqnarray}
% where NNN and RRR are distributions of the pairwise combinations of the galaxies from the data and random catalogs. We calculate the 3PCF results employing a Python toolkit called \texttt{nbodykit}\cite{hand2018nbodykit}\footnote{https://github.com/bccp/nbodykit} designed for use in cosmology. \texttt{nbodykit} adopts \cite{Slepian:2015qza} For the 3pcf calculations. 

% MORE TEXT WITH EQUATIONS HERE

% The details for the calculation can be found in \cite{Slepian:2015qza} and its application to BOSS Data Release 12 \cite{2015ApJS..219...12A} is presented in \citep{Slepian:2015hca}.

% {\bf FIGURES ARE STILL WAITING}

\subsection{Power Spectrum}

%\subsubsection{Spherically Averaged Power Spectrum}
\label{sec:spherically_avg_ps}
To further characterize the large scale structure present in our mock catalogs, we compute the spherically averaged power spectrum. The calculation is performed using \texttt{nbodykit}\cite{hand2018nbodykit}. The process requires mass assignment onto a grid, so it can be sensitive to aliasing effects. We briefly describe the corrections implemented in \texttt{nbodykit}'s \texttt{ConvolvedFFTPower}, which follow the methods of Jing~\cite{jing2005correcting}. 

First, redshifts are converted back to distances according to Equation~\ref{eq:comovingr}). The simulated survey cone is enclosed in a cube, which is then divided into a grid with $450$ points along each coordinate direction, forming a $450^3$ point grid. Values are assigned to the grid points around each galaxy via the Triangle Shaped Cloud\cite{eastwood1981computer} (TSC) window function. A fast Fourier transform (FFT) is applied to the grid according to 
\begin{equation} \label{eq:grid_FFT}
\tilde{n}(\textit{\textbf{k}}) = \sum_{\textit{\textbf{x}}}n(\textit{\textbf{x}})w(\textit{\textbf{x}})\exp({i \textit{\textbf{k}} \cdot \textit{\textbf{x}}}).
\end{equation}
Here, $\textit{\textbf{x}}$ denotes a unique grid point, $n(\textit{\textbf{x}})$ gives the distribution of galaxies over the grid, and $w(\textit{\textbf{x}})$ gives a weighting scheme which is relevant to survey data, but is unity for each galaxy in our mock catalogs. This notation is adopted from Blake\cite{blake2010wigglez}. The power spectrum is then estimated by 
\begin{equation} \label{eq:PS_estimator}
P(\textit{\textbf{k}}) = \frac{|\tilde{n}(\textit{\textbf{k}})-N \tilde{W}(\textit{\textbf{k}})|^2-N\sum_{\textit{\textbf{x}}}W(\textit{\textbf{x}})w^2(\textit{\textbf{x}})}{N^2 N_c \sum_{\textit{\textbf{x}}}W(\textit{\textbf{x}})w^2(\textit{\textbf{x}})}.
\end{equation}
where $N$ is the number of mock galaxies and $N_c$ is the number of cells in the grid and $W(\textit{\textbf{x}})$ is the `survey' selection function which gives the normalized expected galaxy count in each grid cell, and accounts for the shape of the survey relative to the cubic box. This estimate, however, is biased by the original grid assignment. Corrections to this are given at each Fourier wavescale, $\textit{\textbf{k}}$, by
\begin{equation} \label{eq:window_correction}
\frac{\sum_{\textit{\textbf{m}}}H^2(\textit{\textbf{k'}})P(\textit{\textbf{k'}})}{P(\textit{\textbf{k}})}.
\end{equation}
The sum is over vectors of integers, $\textit{\textbf{m}} = (m_x,m_y,m_z)$, where the value of each component can range from zero to the number of grid points in that coordinate direction. $\textit{\textbf{k'}}$ is the wavevector $\textit{\textbf{k}}$, but with $\textit{\textbf{m}}$ multiples of the Nyquist frequency, $\pi n_i / L_i$, added in each direction. $H(\textit{\textbf{k}})$ compensates for the choice of mass assignment, TSC in this case\cite{eastwood1981computer}. It is given in the form
\begin{equation} \label{eq:TSC_window}
H(\textit{\textbf{k}}) = \prod_i [\mathrm{sinc}(\pi k_i / 2 k_N)]^2.
\end{equation}
The power spectra presented here are computed by dividing Eq. \ref{eq:PS_estimator} by Eq. \ref{eq:window_correction}. 

\begin{figure}
    \centering
    \ifdefined\colored
        \includegraphics[width=.47\linewidth]{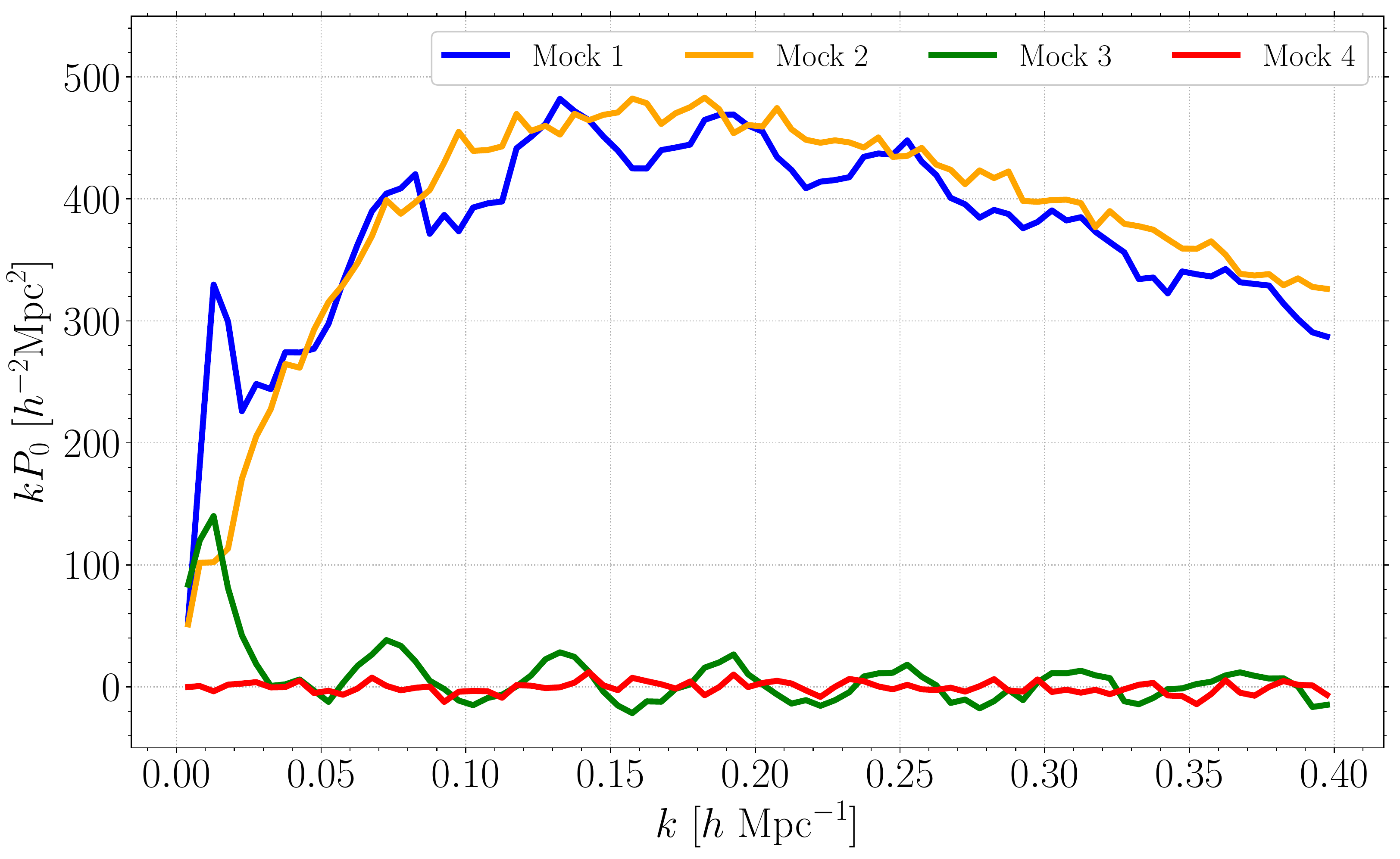}
        \includegraphics[width=.49\linewidth]{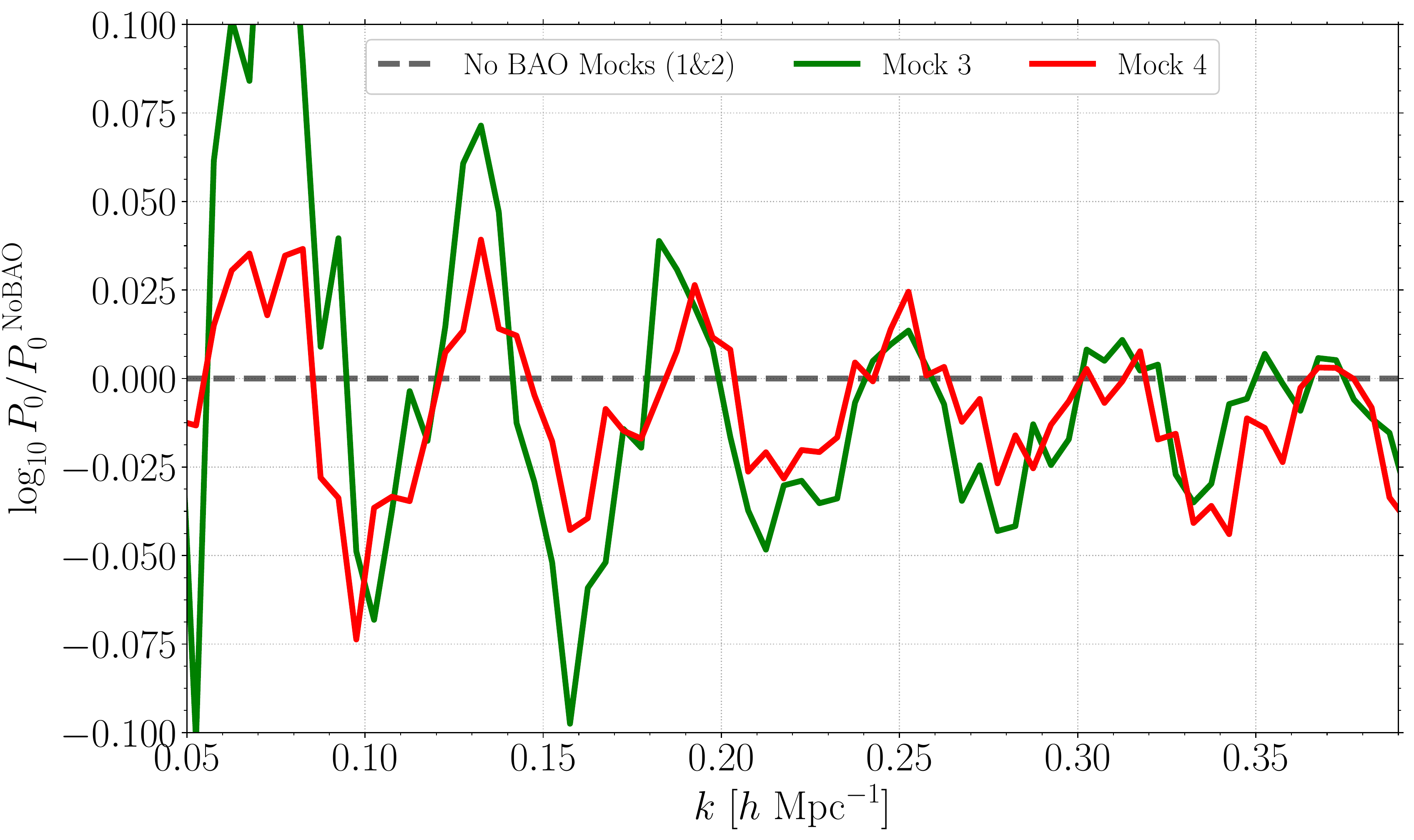}
    \else
        \includegraphics[width=.47\linewidth]{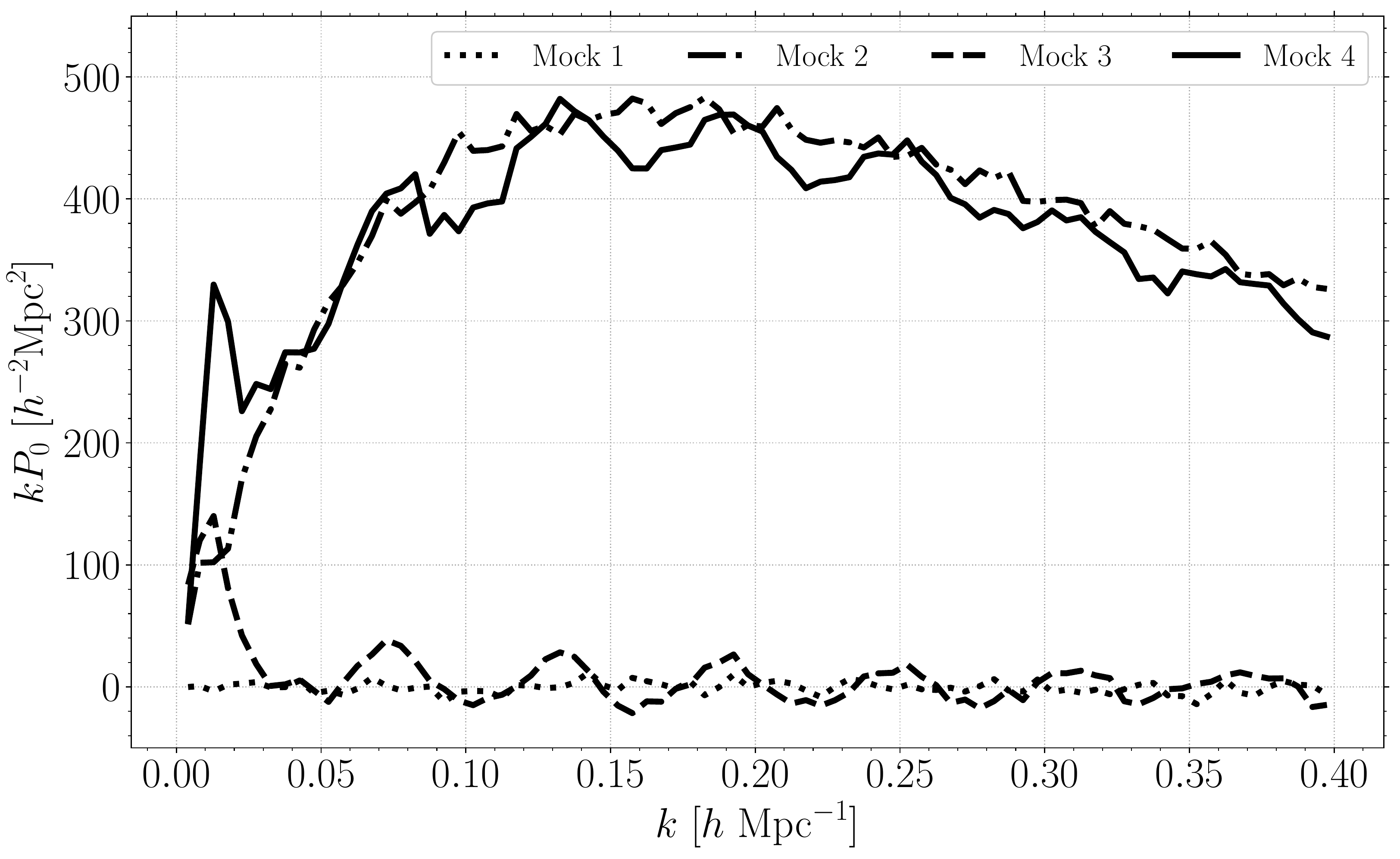}
        \includegraphics[width=.49\linewidth]{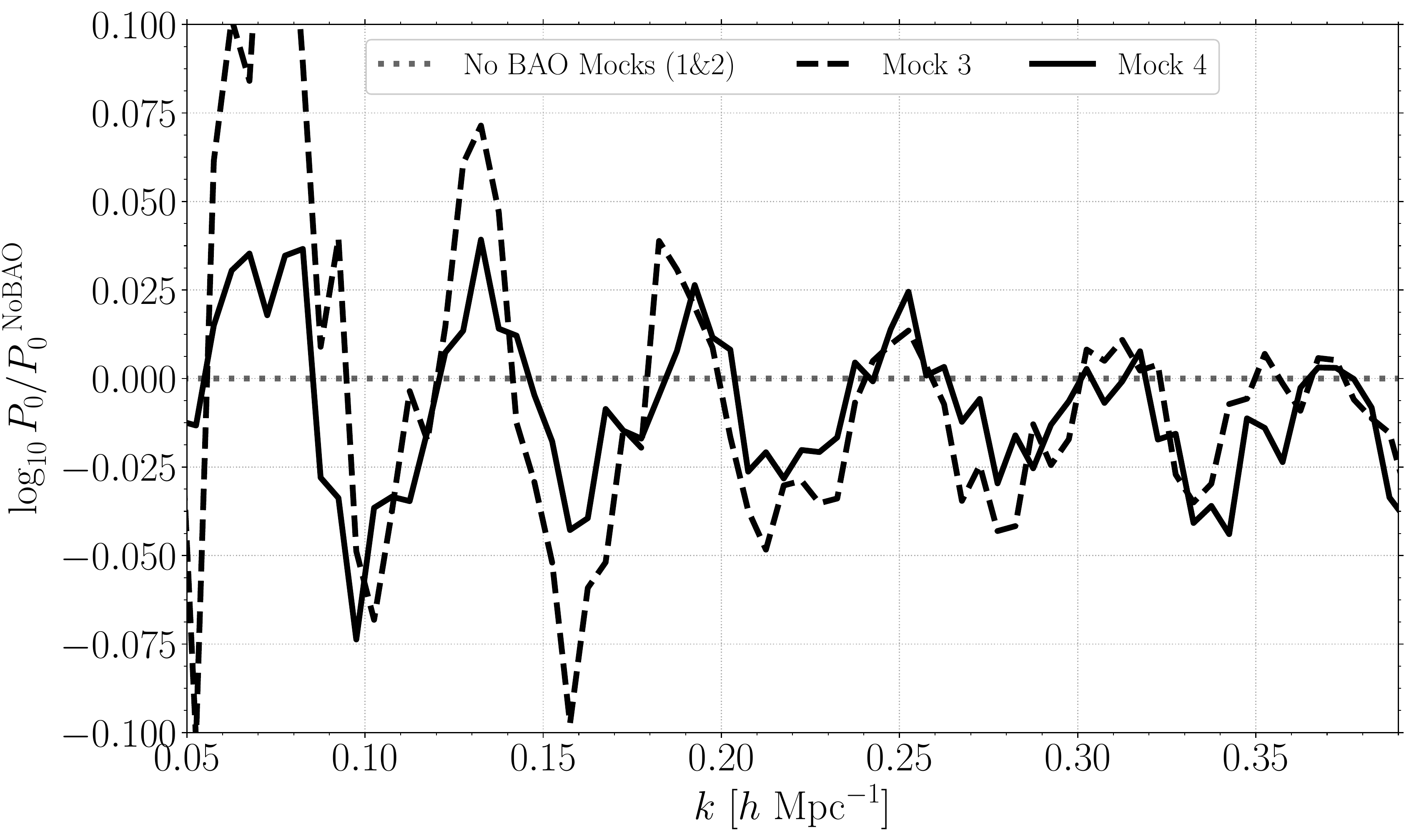}
    \fi
    \caption{\label{fig:power_spectrum} The spherically averaged power spectrum of mocks described in Table \ref{tbl:parameters}, multiplied by the wavescale, $k$ (left). The contribution of BAO structures to the spherically averaged power spectrum, for two of the mocks featured in Table \ref{tbl:parameters} (right). The signal is extracted by comparison to an appropriately similar mock with no center or rim galaxies as described in the text.}
\end{figure}

Figure \ref{fig:power_spectrum} displays the power spectrum of each mock described in Table \ref{tbl:parameters}, multiplied by the wavescale $\textit{\textbf{k}}$. A distinct `wiggling' of the power spectrum is observed in \mockthree and \mockfour. This indicates the presence of a characteristic length scale in real space (the BAO radius). The overall shape of $P(\textit{\textbf{k}})$ for \mocktwo and \mockfour is dominated by the effects of gravitational clumping. This is confirmed by the relatively flat spectra for \mockone and \mockthree which do not contain clumping galaxies. 

Figure \ref{fig:power_spectrum} gives a more detailed picture of the BAO features without the effects of galaxy clumping. The BAO structures' signature oscillations are easily detected by dividing the power spectrum by another reference spectrum and taking the $\log_{10}$. In this case, the reference, or background spectrum for \mockthree and \mockfour is chosen to be \mockone and \mocktwo respectively. For \mockone and \mockthree, 100 $h^{-3}Mpc^3$ are added to each spectrum so that the $\log_{10}$ has a positive argument. To ensure a proper comparison between the magnitude of the signal to the background spectrum, each one is normalized such that the value becomes the probability that a randomly selected density wave contains that wavescale. The strong oscillatory behavior observed in Figure \ref{fig:power_spectrum} confirms the proper modeling of BAO signal in the mock. The periodicity reflects that of a real space length scale around 108 $h^{-1}Mpc$. Furthermore, the extracted BAO signals in \mockthree and \mockfour are similar, reflecting the fact that they were generated with the same values of $N_{center}$ and $n_{rim}$. These power spectra demonstrate the versatility of our simulation, and confirm our expectations about the mocks featured in this study. 

\section{\boldmath Conclusion}
\label{sec:Conclusion}

We have presented a parametric model of large scale structure that allows for a fast evaluation of a wide range of cosmological parameters. This model is realized in an extensive but easy to use Python package, \texttt{ParaMock}\footnote{https://github.com/DESI-UR/catalog\_generation}\cite{yapici_paramock}. The installation instructions and usage are explained in the package. 

The performance of the model was tested on a number of mocks generated using the code. The corresponding 2pcf and power spectra exhibit the features the mocks are generated to model.

The method can be used for a fast evaluation of the galaxy correlations in large spectroscopic surveys. The mocks generated with the parametric model can also be used as a test bench for evaluating the performance of BAO extraction algorithms.

 %In the Appendix, we provide an example configuration file to run

\section*{Acknowledgments}

The authors acknowledge the support from the Department of Energy under the grant DE-SC0008475.

\bibliographystyle{ieeetr}
\bibliography{references}
\label{lastpage}

\end{document}